\documentclass[twocolumn,usenatbib]{aastex61}
\usepackage{amsmath,amssymb}
\received{}
\revised{}
\accepted{}
\submitjournal{ApJ}

\shorttitle{Helium-Star Models with Optically Thick Winds}
\shortauthors{D. Nakauchi and H. Saio}

\begin{document}

\title{Helium-Star Models with Optically Thick Winds: Implications for the Internal Structures and Mass-Loss Rates of Wolf-Rayet Stars}

\author{Daisuke Nakauchi}
\affiliation{Astronomical Institute, Tohoku University, Aoba, Sendai 980-8578, Japan}

\author{Hideyuki Saio}
\affiliation{Astronomical Institute, Tohoku University, Aoba, Sendai 980-8578, Japan}



\begin{abstract}

We construct helium~(He) star models with optically thick winds and compare them with the properties of Galactic Wolf-Rayet~(WR) stars.
Hydrostatic He-core solutions are connected smoothly to trans-sonic wind solutions that satisfy the regularity conditions at the sonic point.
Velocity structures in the supersonic parts are assumed by a simple $\beta$-type law.
By constructing a center-to-surface structure, a mass-loss rate $\dot{M}_{\rm w}$ can be obtained as an eigenvalue of the equations.
Sonic points appear at temperatures $\approx (1.8\mbox{-}2.8) \times 10^{5}\ {\rm K}$ below the Fe-group opacity peak, where the radiation force becomes comparable to the local gravity.
Photospheres are located at radii 3-10 times larger than sonic points.
The obtained mass-loss rates are comparable to those of WR stars.
Our $\dot{M}_{\rm w}$-luminosity relation agrees well with the relation recently obtained by Graefener et al. (2017).
Photospheric temperatures of WR stars tend to be cooler than our predictions.
We discuss the effects of stellar evolution, detailed radiation transfer, and wind clumping, which are ignored in this paper.

\end{abstract}




\section{Introduction}\label{sec:intro}
Wolf-Rayet~(WR) stars are evolved massive stars that show strong and broad emission lines in their spectra.
These features indicate that they have dense envelopes expanding at high speeds, i.e., powerful stellar winds.
The wind mass-loss rates $\dot{M}_{\rm w}$ and terminal velocities $v_\infty$ are observationally determined 
in the range of $\dot{M}_{\rm w} \sim 10^{-5}\mbox{-}10^{-4}\ {\rm M}_\odot\ {\rm yr}^{-1}$ and $v_\infty \sim 1000\mbox{-}3000\ {\rm km}\ {\rm s}^{-1}$,
respectively~\citep{Hamann2006, Sander2012}.

WR stars are classified into two subtypes: WNL/WNE\footnote{
WNL/WNE consists of early~(WNE) and late~(WNL) types.
The former is hydrogen~(H) deficient, while the latter contains a substantial amount of H.
In this paper, we focus only on the early type.} 
showing strong lines of helium~(He) and nitrogen~(N), and 
WC/WO showing lines of He, carbon~(C) and oxygen~(O)~\citep[e.g.,][]{Crowther2007}.
These lines imply that the products of CNO cycle or triple-$\alpha$ reactions 
appear at the surface of WR stars owing to the strong mass loss.
WNE or WC/WO stars are the plausible candidates of the progenitors of Type Ib supernovae~(SNe) 
or Type Ic SNe and even long gamma-ray bursts~\citep[LGRBs;][]{Heger2003, Woosley1993}.

The internal structures of WR stars are often modeled by the evolutionary sequence of He stars, since they have lost most of their hydrogen-rich envelopes by the strong wind mass-loss in the pre-WR phase~\citep[e.g.,][]{Langer1989, McClelland2016,Yoon2017}.
The structure of the wind acceleration region, however, still remains unclear since the sonic point is covered by the dense and optically thick wind.

This is reflected in the so-called radius problem: when the spectroscopically determined radius, $R_\ast$, of a WR star is compared with the core radius of a hydrostatic He zero-age main-sequence~(He-ZAMS) model, the former is larger by up to an order of magnitude~\citep[e.g.,][]{Crowther2007}.
Here, $R_\ast$ usually refers to a radius where the Rosseland-mean optical depth satisfies $\tau_{\rm R}(R_\ast) = 20$, and corresponds to the bottom of an atmosphere model.

To resolve the discrepancy, \citet{Grafener2012} considered hydrostatic He-star models with inflated envelopes by taking account of the iron~(Fe) group opacity peak at $\approx 10^{5.2}\ {\rm K}$~\citep{Iglesias1996} and clumping that increases the inflation of outer envelopes.
Inflated envelopes develop, because an increase in the opacity reduces the local Eddington luminosity and increases the local density scale height.
The inflated tenuous region is surrounded by a thin dense shell caused by the density inversion at an opacity peak.
\citet{McClelland2016} showed that evolved He-star models with clumping have much more extended envelope structures than the He-ZAMS models.

However, envelope inflation may not be compatible with the presence of a powerful wind.
\citet{Petrovic2006} and \citet{McClelland2016} showed that inflated envelope structures disappear from He-star models, if empirical mass-loss rates are included.
Moreover, \citet{Ro2016} calculated expanding envelope solutions and showed that at the high temperature side of the Fe-group opacity peak, the wind is accelerated to a supersonic speed, which erases the structures of density inversion and envelope inflation. 

To understand the internal structures of WR stars from the observed photospheric radii~($r_{\rm ph}$) and temperatures~($T_{\rm ph}$), it is necessary to construct a center-to-surface structure that consists of a hydrodynamic envelope/wind model and an evolutionary model of a hydrostatic core.
\cite{Heger1996} and \cite{Schaerer1996} calculated $r_{\rm ph}$ and $T_{\rm ph}$, by adopting mass-loss rates from the empirical formula derived in \citet{Langer1989b} and a $\beta$-type velocity law:
\begin{equation}
v(r) = v_\infty \left(1-\frac{R_0}{r}\right)^\beta,
\label{eq:beta_law}
\end{equation}
where $R_0$ is the radius at the wind base.
By defining the wind base as a radius where the velocity attains a certain value~(e.g., sound speed), they evaluated the density and temperature at the wind base.
Then these values were used as the outer boundary conditions for the hydrostatic stellar structure equations.
\citet{Groh2013, Groh2014} extended this approach to a wind atmosphere model treating a detailed radiative transfer~\citep{Hillier1998}.
Their models enable to relate the spectral features of massive stars to the internal structures throughout the evolution.
However, the mass-loss rates in their models are assumed by using empirical formulae.

A center-to-surface stellar model constructed by using an assumed mass-loss rate will show discontinuities for the gradients of velocity, density and temperature at the sonic point.
This is because the momentum equation for a spherical and steady trans-sonic flow has a regular singularity at the sonic point and any smooth trans-sonic solution should satisfy the regularity conditions there~\citep[see \S \ref{subsec:match};][]{Kato1992, Lamers1999, Nugis2002, Ro2016}.
A mass-loss rate can be obtained uniquely as an eigenvalue of the equations for the given stellar model, if a center-to-surface stellar model is constructed by combining this regularity conditions at the sonic point with other boundary conditions~(at the stellar center and photosphere)~\citep{Kato1992}.
Based on this idea, \citet{Kato1992} theoretically derived mass-loss rates for their simple He-star models and compared them with the observed WR winds, although they assumed an artificial opacity law in the wind launching region and the supersonic region.

In this paper, we construct a center-to-surface stellar model to study the structures around the wind launching regions and photospheres, and to derive mass-loss rates theoretically.
We have obtained trans-sonic wind solutions by using a method similar to \cite{Kato1992}, but assuming a $\beta$-type velocity law~(Eq. \ref{eq:beta_law}) in the supersonic layers.
We compare the $\dot{M}_{\rm w}$ to luminosity relations and the positions in the Hertzsprung-Russell~(HR) diagram with the observations of Galactic WR stars.
We also compare the $\dot{M}_{\rm w}$ to luminosity relations and the ratio of radiation pressure to gas pressure ($P_{\rm rad}/P_{\rm gas}$) at the sonic point with the results of \cite{Grafener2013} and \cite{Grafener2017}.

The rest of the paper is organized as follows.
In Section \ref{sec:method}, we discuss the basic equations for the hydrostatic core and steady wind, and the boundary conditions.
In Section \ref{sec:result}, we first show the stellar structures of our He-star models.
Then we discuss how the structures in the wind launching regions and photospheres, and mass-loss rates depend on the model parameters.
In Section \ref{sec:obs}, we compare our models with the observations of Galactic WR stars.
The implications and uncertainties of our study are discussed in Section \ref{sec:discussion}.
In Section \ref{sec:summary}, we briefly summarize the results of this paper. 

\section{Models and Methods}\label{sec:method}
Our He-star models consist of two regions: a hydrostatic core and steady wind.
The latter is further divided into the inner subsonic and the outer supersonic regions.

\subsection{Hydrostatic He-Core}\label{subsec:core}
The basic equations to construct a hydrostatic He-burning core are as follows~\citep[e.g.,][]{Kippenhahn2012}:
\begin{equation}
\frac{d r}{d M_r} = \frac{1}{4 \pi r^2 \rho},
\label{eq:star_eoc}
\end{equation}
\begin{equation}
\frac{d P}{d M_r} = - \frac{G M_r}{4 \pi r^4},
\label{eq:star_eom}
\end{equation}
\begin{equation}
\frac{d L_r}{d M_r} = \epsilon_{\rm nuc},
\label{eq:star_eoe}
\end{equation}
\begin{equation}
L_r = L_{\rm rad} + L_{\rm conv},
\label{eq:ene_transfer}
\end{equation}
where $G$ is the gravitational constant, $P$ the total pressure, $M_r$ the enclosed mass within the radius $r$, $\rho$ the density, $L_r$ the total luminosity, $\epsilon_{\rm nuc}$ the nuclear energy generation rate via the triple-$\alpha$ reaction~(3He$^4 \rightarrow$ C$^{12}$; Eq. 5-104 of \citet{Clayton1983}), $L_{\rm rad}$ the radiative luminosity, and $L_{\rm conv}$ the convective luminosity.
The total pressure $P$ is composed of the radiation pressure $P_{\rm rad}$ and the gas pressure $P_{\rm gas}$:
\begin{equation}
P = P_{\rm gas} + P_{\rm rad} = \frac{\mathcal{R}}{\mu} \rho T + \frac{1}{3} a_{\rm rad} T^4,
\label{eq:eos}
\end{equation}
where $a_{\rm rad}$ is the radiation constant, $\mathcal{R}$ the gas constant, $T$ the temperature, and $\mu$ the mean molecular weight.
Since we consider chemically homogeneous He stars, $\mu$ is taken as a constant throughout
the core and wind in this paper.
Radiative luminosity $L_{\rm rad}$ is calculated using the diffusion approximation:
\begin{equation}
L_{\rm rad} = - \frac{16 \pi a_{\rm rad} c r^2 T^3}{3 \kappa_{\rm R} \rho}\frac{dT}{dr},
\label{eq:l_rad}
\end{equation}
where $c$ is the speed of light, and $\kappa_{\rm R}(\rho, T)$ the Rosseland mean opacity.
The opacity, $\kappa_{\rm R}$, is obtained from the tables provided by the OPAL project~\citep{Iglesias1996}, and is calculated by using the bilinear interpolation.
Convective luminosity $L_{\rm conv}$ is calculated by using the mixing-length theory~\citep{Eggleton1971},
if the radiative temperature gradient, $\nabla_{\rm rad}$, is larger than the adiabatic one, $\nabla_{\rm ad}$~\citep[the Schwarzschild criterion, e.g.,][]{Kippenhahn2012}:
\begin{equation}
\nabla_{\rm rad} \geq \nabla_{\rm ad},
\label{eq:schwarzschild}
\end{equation}
where $\nabla_{\rm ad} \equiv (d \log T / d \log P)_{\rm ad} = 2 (4 - 3 \beta_P) / (32 - 24 \beta_P - 3 \beta_P^2)$ with
$\beta_P \equiv P_{\rm gas}/P$ and
\begin{equation}
\nabla_{\rm rad} = \frac{d \log T}{d \log P} = \frac{3 \kappa_{\rm R} L_r P}{16 \pi a_{\rm rad} c G M_r T^4},
\label{eq:nab_rad_core}
\end{equation}
in the core.
Otherwise, all of the luminosity can be transported by radiation, so that $L_{\rm r} = L_{\rm rad}$ and $L_{\rm conv} = 0$.

\subsection{Steady Wind Model}\label{subsec:wind}
The structure of a steady wind is calculated from the following equations. 
Note that our formulation is valid as long as the wind is optically thick~\citep[e.g.,][]{Zytkow1972,Kato1992,Nakauchi2017}.
\begin{equation}
\dot{M}_{\rm w} \equiv 4 \pi r^2 \rho v = {\rm const.},
\label{eq:eoc}
\end{equation}
\begin{equation}
\frac{d M_r}{dr} = 4 \pi r^2 \rho,
\label{eq:eoc2}
\end{equation}
\begin{equation}
v \frac{dv}{dr} + \frac{1}{\rho} \frac{dP}{dr} + \frac{GM_r}{r^2} = 0,
\label{eq:eom}
\end{equation}
\begin{equation}
\Lambda \equiv L_{\rm r} + \dot{M}_{\rm w} \left(\frac{v^2}{2} + \frac{5 \mathcal{R} T}{2 \mu} + \frac{4 a_{\rm rad} T^4}{3 \rho} + \int_{r_{\rm s}}^{r} \frac{GM_r}{r^2} dr\right) = {\rm const.},
\label{eq:energy}
\end{equation}
where $v$ is the wind velocity, $\dot{M}_{\rm w}$ the mass-loss rate, $\Lambda$ the energy constant, $r_{\rm s}$ the sonic radius.
In Eq. \eqref{eq:energy}, we neglect the nuclear energy generation in the wind region, since the wind temperature is too low for He-burning to be significant.
It should be noted that in the limit of $\dot{M}_{\rm w} \rightarrow 0$ and $v dv/dr \rightarrow 0$, Eqs.~(\ref{eq:eoc}-\ref{eq:energy})
return to Eqs.~(\ref{eq:star_eoc}-\ref{eq:star_eoe}) with $\epsilon_{\rm nuc} = 0$.

In the inner subsonic region, the total luminosity $L_{\rm r}$, the radiative luminosity $L_{\rm rad}$, and the convective luminosity $L_{\rm conv}$ are calculated from Eqs. \eqref{eq:ene_transfer} and \eqref{eq:l_rad}, and by using the mixing length theory, respectively.
However, by using Eq. \eqref{eq:eom}, $\nabla_{\rm rad}$ is calculated without assuming the hydrostatic equilibrium as
\begin{equation}
\nabla_{\rm rad} = \frac{3 \kappa_{\rm R} L_r P}{16 \pi a_{\rm rad} c r^2 T^4} \left(- \frac{1}{\rho} \frac{dP}{dr}\right)^{-1}.
\label{eq:nab_rad_wind}
\end{equation}

In the outer supersonic region, instead of calculating hydrodynamical equations coupled with the radiative transfer equation~\citep{Grafener2005, Grafener2008, Sander2017}, we assume a $\beta$-type velocity law~(Eq. \ref{eq:beta_law}).
In the supersonic layers~(where $P_{\rm rad} \gg  P_{\rm gas}$ holds), this assumption is nearly equivalent to replace $\kappa_{\rm R}$ with the effective opacity represented by
\begin{equation}
\kappa_{\rm eff}(r) = \kappa_{\rm R}(\rho_{\rm s}, T_{\rm s}) \left[1 + 2 \beta \left(\frac{v_\infty}{v_{\rm esc}(r_{\rm s})}\right)^2 \left(1-\frac{r_{\rm s}}{r}\right)^{2 \beta -1} \right],
\label{eq:kap_eff}
\end{equation}
where $\kappa_{\rm R}(\rho_{\rm s}, T_{\rm s})$ and $v_{\rm esc}(r_{\rm s})$ are the Rosseland-mean opacity and the escape velocity at the sonic radius, respectively~\citep{Lucy1993,Grafener2013}.
If we solve Eqs.~(\ref{eq:eoc}-\ref{eq:energy}) with $L_{\rm rad}$~(Eq. \ref{eq:l_rad}) by replacing $\kappa_{\rm R}$ with $\kappa_{\rm eff}$, we obtain a $\beta$-type velocity law~(Eq. \ref{eq:beta_law}) with $R_0 = r_{\rm s}$.
Note that in the supersonic region, we neglect the convective energy transport in Eq. \eqref{eq:ene_transfer}, i.e., $L_{\rm r} = L_{\rm rad}$, since the velocity of the convective element should be less than the sound speed and it may not exceed the energy transport by advection.

The effective opacity~(Eq. \ref{eq:kap_eff}) is characterized by two parameters: $v_\infty/v_{\rm esc}(r_{\rm s})$ and $\beta$.
According to the non-local thermodynamic equilibrium~(LTE) wind models of \cite{Grafener2005}, the velocity structure can be approximately fitted with $\beta = 1$ in the inner rapidly accelerating region.
Moreover, $\beta \approx 0.7-0.8$ is obtained from the observations of the O star winds~\citep{Puls2008}.
On the other hand, \cite{Grafener2013}'s models indicate $v_\infty/v_{\rm esc}(r_{\rm s}) \approx 1.6$ for WC and WO winds.
For O star winds, it is known that $v_\infty/v_{\rm esc} \approx \mathcal{O}(1)$~\citep[e.g.,][]{Kudritzki2000}.
Therefore, we study the cases with $v_\infty/v_{\rm esc}(r_{\rm s}) \approx 1\mbox{-}2$ and $\beta = 0.75$ and $1$, respectively.

\subsection{Connecting Steady Wind to Hydrostatic Core}\label{subsec:match}
In the steady wind model, equations~(\ref{eq:ene_transfer}, \ref{eq:eoc}-\ref{eq:energy}) have five unknown functions,
$v(r), \rho(r), T(r), L_r$, and $M_r$, while in the core region, equations~(\ref{eq:star_eoc}-\ref{eq:ene_transfer}) have four unknowns~(since $v(r) = 0$).
A wind solution that is smoothly connected to a core solution can be obtained by providing five boundary conditions:
two of them at the core center, other two at the sonic point, and the last one at the photosphere~\citep{Kato1992}.

First, at the core center, the luminosity and enclosed mass should become zero:
\begin{equation}
L_r = 0\ \text{and}\ M_r = 0\ \text{at}\ r = 0.
\label{eq:bc_center}
\end{equation}
Therefore, we can obtain one core solution, if we give the density and temperature at the core center, $\rho_{\rm c}$ and $T_{\rm c}$.
 
Second, the following regularity conditions should be satisfied at the sonic point.
By substituting Eqs. \eqref{eq:eos} and \eqref{eq:eoc} into Eq. \eqref{eq:eom}, it is rewritten as
\begin{equation}
\frac{1}{v} \frac{d v}{d r} = \left[\frac{2}{r}c_{\rm T}^2 - \frac{1}{\rho} \left(\frac{\partial P}{\partial T}\right)_\rho \frac{d T}{dr} - \frac{GM_r}{r^2} \right] / \left(v^2 - c_{\rm T}^2 \right),
\label{eq:vel_grad}
\end{equation}
where $c_{\rm T} = \sqrt{(\partial P/\partial \rho)_T}$ is the isothermal sound speed.
We can see from Eq. \eqref{eq:vel_grad} that the sonic point is the singular point of the equation.
A transonic wind solution can be obtained by requiring that the numerator of the equation vanishes at the sonic point
and that the velocity gradient become finite there~\citep[e.g.,][]{Lamers1999}.
From these regularity conditions, we can evaluate $dT/dr$~(and hence the radiative luminosity) and the wind velocity at the sonic point,
for given values of the radius, density, and temperature there, $r_{\rm s}, \rho_{\rm s}$, and $T_{\rm s}$.
We also equate the enclosed mass there with the total mass of a star, $M_\ast$, since the mass within the supersonic region
must be much less than $M_\ast$.
To summarize, the following two boundary conditions are imposed at the sonic point:
\begin{eqnarray}
v(r_{\rm s}) &=& c_{\rm T}(\rho_{\rm s}, T_{\rm s}), \\
L_{\rm rad}(r_{\rm s}) &=& L_{\rm rad}(r_{\rm s}, \rho_{\rm s}, T_{\rm s})\ \text{at}\ M(r_{\rm s}) = M_\ast. \notag
\label{eq:bc_sonic}
\end{eqnarray}
It should be noted that once $r_{\rm s}, \rho_{\rm s}$, and $T_{\rm s}$ are specified, we can obtain one wind solution.
This is because, for a given set of $(r_{\rm s}, \rho_{\rm s}, T_{\rm s})$, $\dot{M}_{\rm w}$ and $\Lambda$ can be evaluated from Eqs. \eqref{eq:eoc} and \eqref{eq:energy}, respectively, and the velocity gradient at the sonic point by using the de l'Hopital rule to Eq. \eqref{eq:vel_grad}~\citep{Lamers1999, Nugis2002}.

Finally, at the photospheric radius~($r_{\rm ph}$), where the effective temperature $T_{\rm eff}(r_{\rm ph})$ becomes equal to the local temperature, we require that the optical-depth-like variable $\tau(r) \equiv \kappa_{\rm eff}(r) \rho(r) r$ becomes equal to $8/3 \approx 2.7$~\citep{Kato1992, Kato1994}.
Therefore, the boundary condition at the photosphere are represented as
\begin{equation}
T_{\rm eff}(r_{\rm ph}) = T(r_{\rm ph}) \equiv T_{\rm ph}\ \text{and}\ \tau(r_{\rm ph}) = 2.7.
\label{eq:bc_photo}
\end{equation}
By using the effective opacity $\kappa_{\rm eff}$, we perform the outward integration from the sonic radius and find solutions that satisfy the photospheric condition.

The values of the five parameters, $\rho_{\rm c}$, $T_{\rm c}$, $r_{\rm s}$, $\rho_{\rm s}$, and $T_{\rm s}$, are determined iteratively
so that the wind solution is smoothly connected to the hydrostatic core solution at some radius $r_{\rm m}$ between the core center and the sonic point, satisfying $T(r_{\rm m}) \simeq 10^8\ {\rm K}$.
Therefore, the energy generation by He-burning is negligible for $r \geq r_{\rm m}$.
We first fix the value of $\rho_{\rm s}$, and then iteratively determine the values of $\rho_{\rm c}$, $T_{\rm c}$, $r_{\rm s}$, and $T_{\rm s}$
so that the enclosed mass, density, and temperature should be continuous, and the total luminosity should be conserved
at the matching radius $r_{\rm m}$:
\begin{eqnarray}
M_{\rm core}(r_{\rm m}) &=& M_{\rm wind}(r_{\rm m}),\notag \\
\rho_{\rm core}(r_{\rm m}) &=& \rho_{\rm wind}(r_{\rm m}),\\
T_{\rm core}(r_{\rm m}) &=& T_{\rm wind}(r_{\rm m}), \notag
\label{eq:bc_rho_temp}
\end{eqnarray}
and
\begin{eqnarray}
\label{eq:bc_energy_flux}
L_{\rm core}(r_{\rm m}) &=& L_{\rm wind}(r_{\rm m}) \notag \\
&+& \dot{M}_{\rm w} \left(\frac{v^2(r_{\rm m})}{2} + \frac{5 \mathcal{R}}{2 \mu} T_{\rm wind}(r_{\rm m}) + \frac{4 a_{\rm rad}}{3} \frac{T_{\rm wind}^4(r_{\rm m})}{\rho_{\rm wind}(r_{\rm m})} \right) \notag \\
&=& \Lambda + \dot{M}_{\rm w} \int_{r_{\rm m}}^{r_{\rm s}} \frac{GM_r}{r^2} dr.
\end{eqnarray}
Next, $\rho_{\rm s}$ is determined so that the wind solution satisfies the photospheric condition Eq. \eqref{eq:bc_photo}.
Then, we obtain a He-star model with an optically thick wind.

\subsection{Chemical Composition}\label{subsec:chem}
In this paper, we consider two types of He-star models, {\it He-rich} and {\it CO-enriched} models.
Since we consider H-free stellar models, the mass fraction of H is set to zero, $X = 0$, in both models.
The mass fraction of He, $Y$, is calculated from $Y = 1- \tilde{Z}$, where $\tilde{Z}$ is the mass fraction of metals heavier than He.
In He-rich models, we suppose that $\tilde{Z}$ is identical with the metallicity $Z$ in the solar composition~\citep{Grevesse1993},
while in CO-enriched models, the mass fractions of C and O are enhanced by $dX_{\rm C} = 0.4$ and $dX_{\rm O} = 0.1$, i.e., $\tilde{Z} = Z + dX_{\rm C} + dX_{\rm O}$.
In each model, we consider two metallicities of $Z = 1$ and $2\ {\rm Z}_\odot$ with ${\rm Z}_\odot = 0.02$~(solar metallicity)\footnote{The exact value of the solar metallicity ${\rm Z}_\odot$ is still uncertain. We adopt ${\rm Z}_\odot = 0.02$ as the solar metallicity, since the adopted opacity table is based on the solar composition in \cite{Grevesse1993}.}.

Our He-star models may be too simple compared to actual WR stars.
Especially, WC stars may have compositional gradients between CO cores and He-rich envelopes.
We note, however, that our purpose in this paper is to construct whole-star models with optically thick winds, in which a subsonic wind structure is connected smoothly to a supersonic part at the sonic point.
This enables us to determine a mass-loss rate uniquely for a given mass, chemical composition, and opacity parameters.
Our simple models are enough to understand the dependences of the structures around the wind launching regions and photospheres, and of mass-loss rates on these parameters.

\section{Results}\label{sec:result}

\subsection{He-rich Models}\label{subsec:WN}
\subsubsection{Stellar Structure}\label{subsec:WN_structure}
\begin{figure*}
\begin{center}
\includegraphics[scale=0.6, angle=-90]{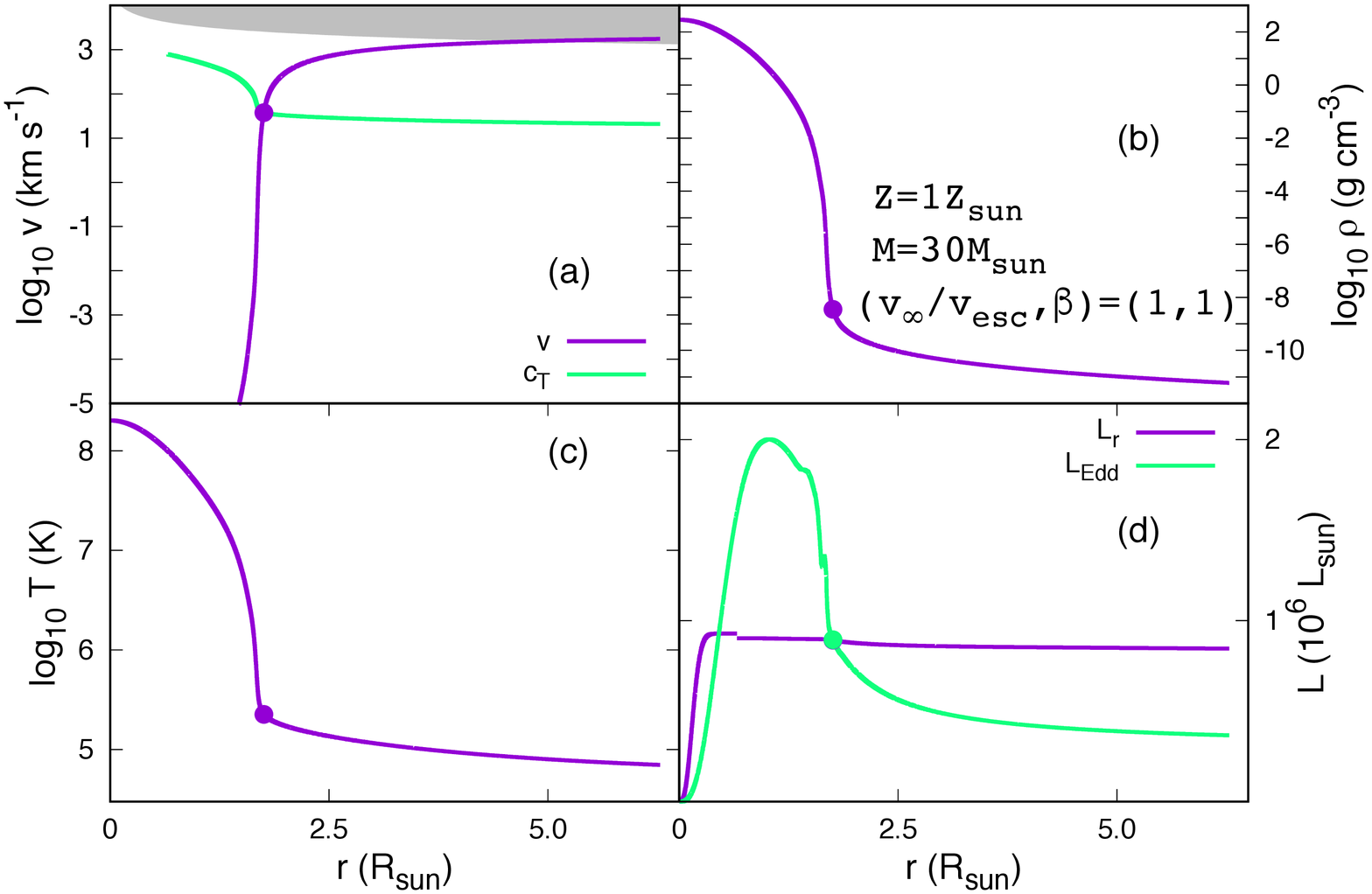}
\caption{Stellar structures of a He-rich model with $M_\ast = 30\ {\rm M}_\odot$ and $Z = 1\ {\rm Z}_\odot$.
Opacity parameters are adopted as $(v_\infty/v_{\rm esc}(r_{\rm s}), \beta) = (1.0, 1.0)$.
In each panel, the filled circle shows the location of the sonic point.
{\it Panel} a): The velocity structure of the wind. 
The purple line shows the wind velocity, the green line isothermal sound speed, and the grey-shaded region
where $v(r) \geq v_{\rm esc}(r)$ holds. 
{\it Panel} b): The density profile.
{\it Panel} c): The temperature profile.
{\it Panel} d): The luminosity structure. The purple line shows the total luminosity and the green one the local Eddington luminosity. 
For this model, the mass-loss rate is obtained as $\dot{M}_{\rm w} \simeq 3.98 \times 10^{-5}\ {\rm M}_\odot\ {\rm yr}^{-1}$.}
\label{fig:wn_m30_z1_a20_b10}
\end{center}
\end{figure*}

First, we show the stellar structure of a He-rich model with $M_\ast = 30\ {\rm M}_\odot$ and $Z = 1\ {\rm Z}_\odot$~(Fig. \ref{fig:wn_m30_z1_a20_b10}).
Opacity parameters are adopted as $(v_\infty/v_{\rm esc}(r_{\rm s}), \beta) = (1.0, 1.0)$.
In this model, the matching point to the static core is located at $r_{\rm m} \simeq 0.660\ {\rm R}_\odot$, the sonic point~(filled circle) at $r_{\rm s} \simeq 1.76\ {\rm R}_\odot$, and the photospheric radius at $r_{\rm ph} \simeq 6.29\ {\rm R}_\odot$.
While the wind velocity is sufficiently subsonic~($v \lesssim 1\ {\rm cm}\ {\rm s}^{-1}$) around $r_{\rm m}$,
it is steeply accelerated to a supersonic speed passing through the sonic radius $r_{\rm s}$~(panel a).
It finally reaches a constant speed of $v_\infty \sim 1750\ {\rm km}\ {\rm s}^{-1}$, which is larger than the escape velocity at the photosphere $r_{\rm ph}$.
The photospheric radius $r_{\rm ph}$ is several times larger than the sonic radius $r_{\rm s}$. 
In this model, the mass-loss rate is evaluated as $\dot{M}_{\rm w} \simeq 3.98 \times 10^{-5}\ {\rm M}_\odot\ {\rm yr}^{-1}$. 
These values are within the range of the observed WNE stars~\citep{Hamann2006}.

The density and temperature decline outward steeply to the sonic radius $r_{\rm s}$ in subsonic layers~(see panels b and c of Fig. \ref{fig:wn_m30_z1_a20_b10}).
In the highly supersonic region, the density is inversely proportional to the square of radius, $\rho \propto r^{-2}$,
since the wind velocity is almost constant there.
The integration is terminated at the photosphere $r_{\rm ph}$ with $\log T_{\rm ph} \simeq 4.85$. 
We also find that the radiation pressure dominates the gas pressure over the supersonic region, while they are almost the same order of magnitude in the hydrostatic core and deep subsonic part.

The luminosity generated via the nuclear burning in the core, $L_{\rm core} \simeq 9.29 \times 10^5\ {\rm L}_\odot$~(panel d of Fig. \ref{fig:wn_m30_z1_a20_b10}), is consistent with the mass$-$luminosity relation obtained from the hydrostatic He-star models~\citep[Eq. 3 of][]{Schaerer1992}, which gives $L_{\rm hs} \simeq 1.02 \times 10^6\ {\rm L}_\odot$ for $M_\ast = 30\ M_\odot$.
At $r = r_{\rm m}$, a portion of the core luminosity is converted into the mechanical luminosity of the wind,
which causes a tiny discontinuity in the local luminosity there.
From there to the surface, the local luminosity slightly decreases to the photospheric value,
$L_{\rm ph} \simeq 8.45 \times 10^5\ {\rm L}_\odot$.
This indicates that only a small fraction~($\simeq 9\ \%$) of the core luminosity is consumed for accelerating the wind.

The sonic point is located at the radius where the luminosity becomes equal to the local Eddington luminosity~(panel d), i.e, the radiation force becomes equal to the local gravity.
Beyond the sonic point, the luminosity should be super-Eddington, in order for the wind velocity to be accelerated up to the terminal value.
Note that the sonic radius $r_{\rm s} \simeq 1.76\ {\rm R}_\odot$ is only slightly larger than the radius of a hydrostatic He-star model~($R_{\rm hs} \simeq 1.58\ {\rm R}_\odot$) calculated from Eq. 4 of \cite{Schaerer1992}.

\begin{figure}
\begin{center}
\includegraphics[scale=0.3, angle=-90]{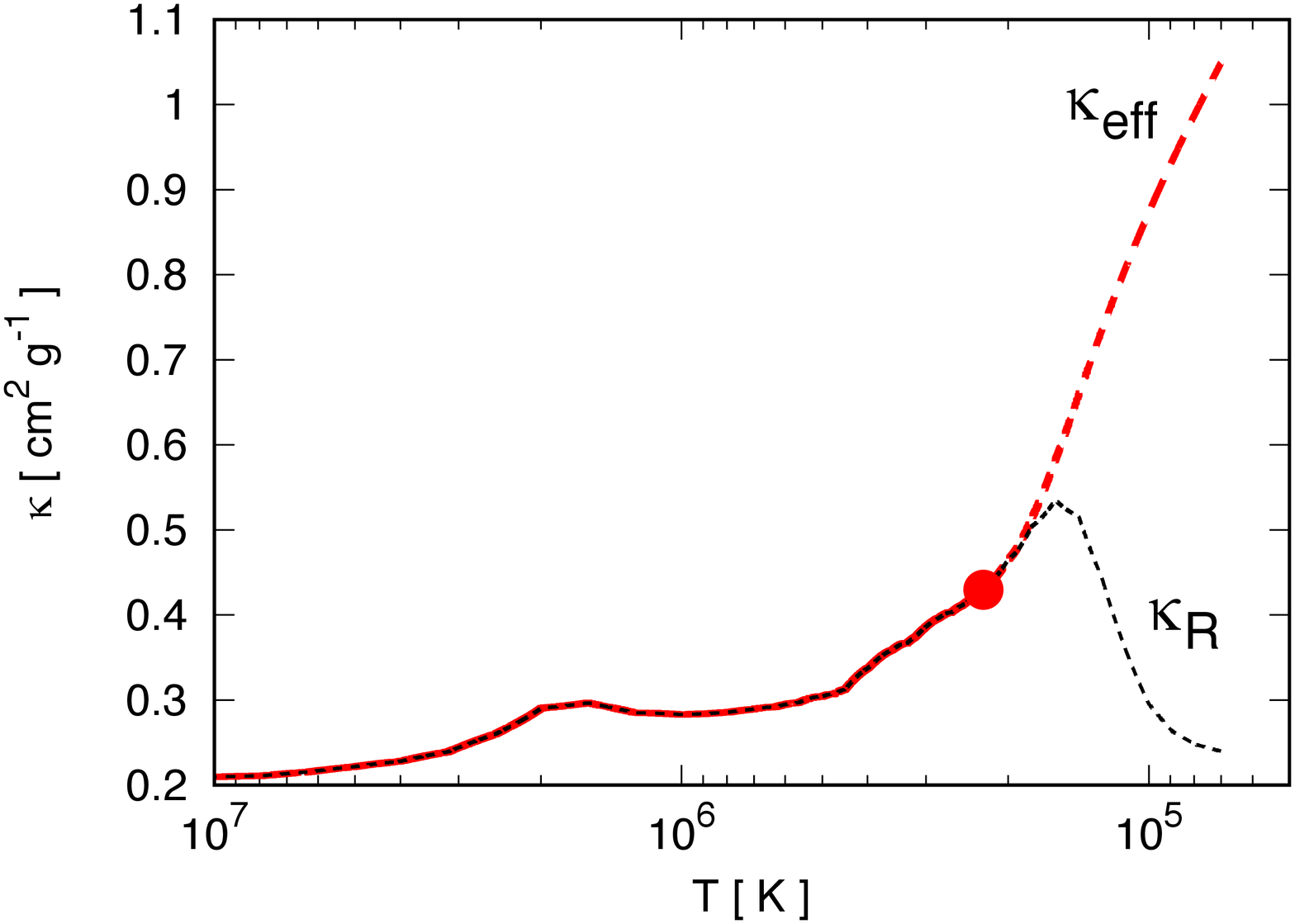}
\caption{The opacity structure in the He-rich model of Fig. \ref{fig:wn_m30_z1_a20_b10}.
The filled circle indicates the location of the sonic point. 
Above the sonic point~(red dashed line), the opacity is calculated from the semi-analytical formula in Eq. \eqref{eq:kap_eff}.
For comparison, the Rosseland-mean opacity calculated from the OPAL table is shown with the thin black dashed line.
}
\label{fig:wn_temp_kap}
\end{center}
\end{figure}

Fig. \ref{fig:wn_temp_kap} shows the opacity as a function of temperature in the wind.
The filled circle indicates the location of the sonic point.
In the subsonic part, we adopt the Rosseland mean opacity $\kappa_{\rm R}$ from the OPAL table~(the thin black dashed line).
The rapid acceleration of the subsonic layers starts with the increase of the Fe-group opacity, and the wind velocity reaches the sound speed at $T_{\rm s} \simeq 2.26 \times 10^5\ {\rm K}$, before the opacity reaches the peak at $\approx 1.5 \times 10^5\ {\rm K}$.
The importance of the Fe opacity for wind ignition is also found in the previous studies~\citep{Nugis2002, Grafener2005}.
Beyond the sonic point, the opacity is evaluated from the semi-analytical formula in Eq. \eqref{eq:kap_eff}~($\kappa_{\rm eff}$, red dashed line).
This implies that in order to realize the assumed $\beta$-type velocity profile, the effective opacity must be increased by more than a factor of a few compared to the Rosseland mean opacity $\kappa_{\rm R}$ calculated in a static medium.
The opacity enhancement in supersonic layers can be expected from the effects of the non-LTE, Doppler shifting of spectral lines, and wind clumping.

\subsubsection{Parameter Dependences}\label{subsec:WN_mdot}

\begin{table}
\caption{Summary for the symbols, line types, and colors used in Figs. 3-5 and 7-11.}
   \centering
   \begin{tabular}{lccc}
      \hline
      Opacity parameters: & \multicolumn{3}{c}{($v_\infty/v_{\rm esc}(r_{\rm s}),~\beta$ )} \\
      \hline
      Symbol:& $\bigcirc$   & $\bigtriangleup$  & $\Box$ \\
      Line type:& solid & dashed & dot-dashed \\
      \hline
      Red ($1\ {\rm Z}_\odot$): & $(2.3,0.75)$    & $(2.0,1.0)$ &  $(1.0,1.0)$ \\
      Blue ($2\ {\rm Z}_\odot$): &  $(1.6,0.75)$    & $(1.4,1.0)$ &  $(1.0,1.0)$ \\
     \hline
   \end{tabular}
   \label{tab:parameter}
\end{table}

We discuss, in this subsection, the properties of He-rich models obtained for various stellar masses~($M_\ast = 10-60\ {\rm M}_\odot$), metallicities~($Z = 1, 2\ {\rm Z}_\odot$), and opacity parameters~$(v_\infty/v_{\rm esc}(r_{\rm s}), \beta)$.
In the following figures, red and blue colors show the models with $Z = 1\ {\rm Z}_\odot$ and $2\ {\rm Z}_\odot$, respectively.
The open circles connected by solid lines, open triangles by dashed lines, and open squares by dash-dotted lines correspond to the models with the different opacity parameters: $(v_\infty/v_{\rm esc}(r_{\rm s}), \beta) = (2.3, 0.75), (2.0, 1.0)$, and $(1.0, 1.0)$ for $Z = 1\ {\rm Z}_\odot$ case, and $(v_\infty/v_{\rm esc}(r_{\rm s}), \beta) = (1.6, 0.75), (1.4, 1.0)$, and $(1.0, 1.0)$ for $Z = 2\ {\rm Z}_\odot$ case, respectively.
These are summarized in Table \ref{tab:parameter}.

\begin{figure}
\begin{center}
\begin{tabular}{c}
{\includegraphics[scale=0.3, angle=-90]{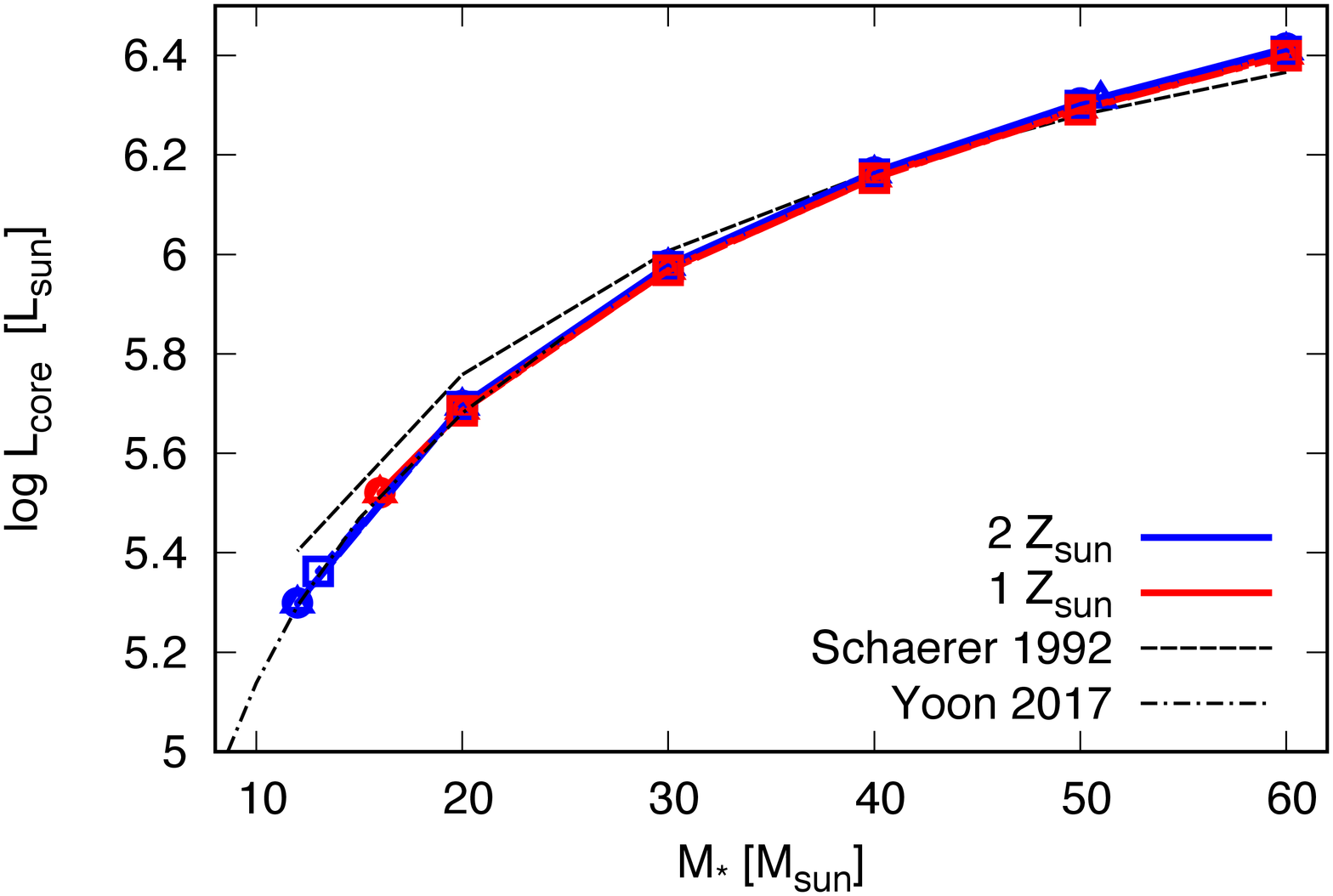}}\\
{\includegraphics[scale=0.3, angle=-90]{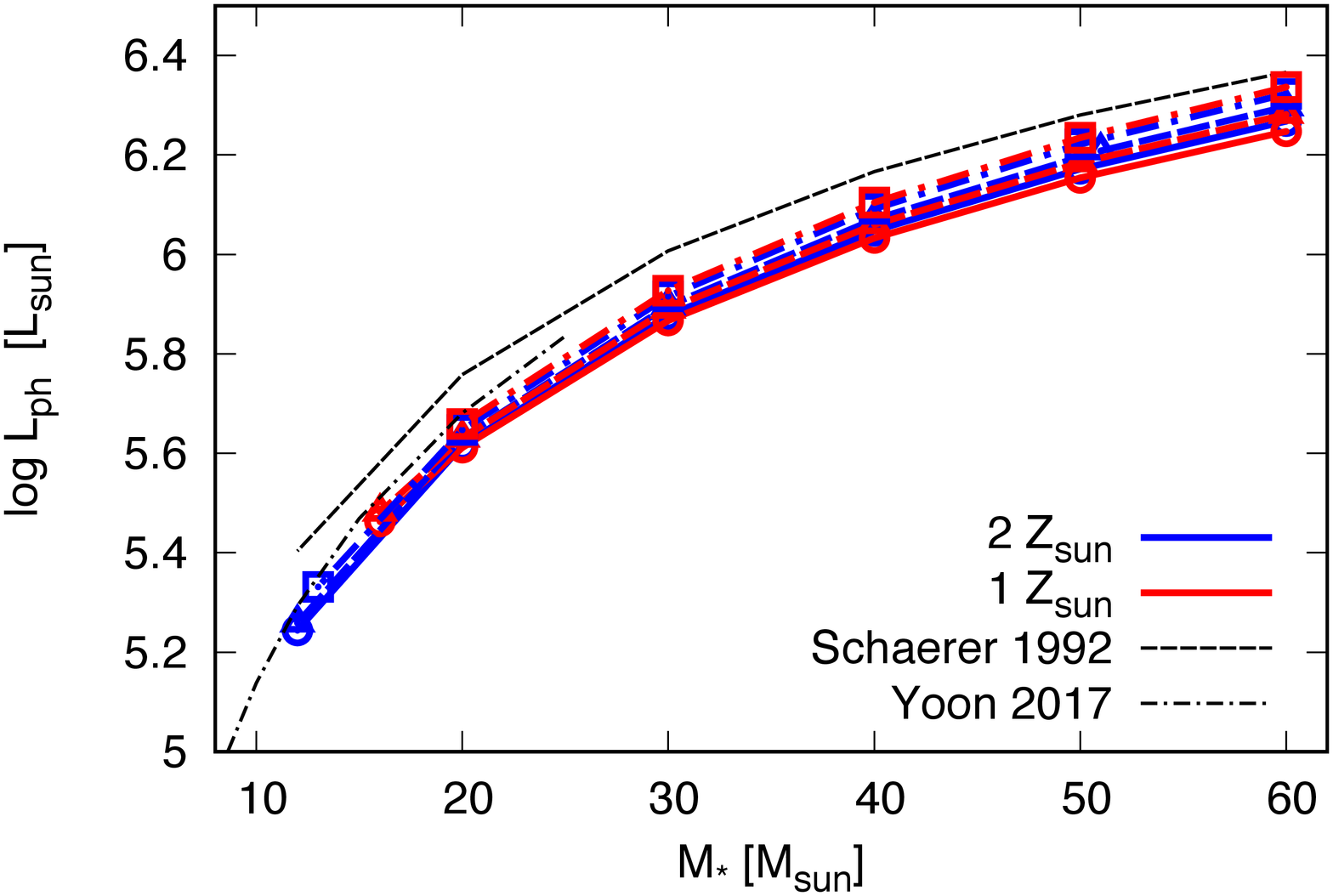}}\\
\end{tabular}
\caption{The core luminosity generated via the nuclear burning~(top panel) and the luminosity at the photosphere~(bottom panel) as a function of the stellar mass.
Red and blue colors show the models with $Z = 1\ {\rm Z}_\odot$ and $2\ {\rm Z}_\odot$, respectively.
The open circles connected by solid lines, open triangles by dashed lines, and open squares by dash-dotted lines correspond to the models with the different opacity parameters: $(v_\infty/v_{\rm esc}(r_{\rm s}), \beta) = (2.3, 0.75), (2.0, 1.0)$, and $(1.0, 1.0)$ for $Z = 1\ {\rm Z}_\odot$ case, and $(v_\infty/v_{\rm esc}(r_{\rm s}), \beta) = (1.6, 0.75), (1.4, 1.0)$, and $(1.0, 1.0)$ for $Z = 2\ {\rm Z}_\odot$ case, respectively~(see also, Table \ref{tab:parameter}).
The thin black dashed and dash-dotted lines show the mass-luminosity relations of the hydrostatic He-star models calculated by \cite{Schaerer1992}~(their Eq. 3) and by \cite{Yoon2017}, respectively.} 
\label{fig:wn_lum}
\end{center}
\end{figure}

In Fig. \ref{fig:wn_lum}, the top and bottom panels show the core luminosity generated via the nuclear burning and the luminosity at the photosphere as a function of the stellar mass, respectively.
The thin black dashed and dash-dotted lines show the mass-luminosity relations of the hydrostatic He-star models calculated by \cite{Schaerer1992}~(their Eq. 3) and by \cite{Yoon2017}, respectively.
Since the core luminosity is determined thoroughly by the core structure, it is independent of the opacity parameters~(the top panel).
The mass to core luminosity relation agrees well with those of \cite{Schaerer1992} and \cite{Yoon2017}, meaning that the core structures of our He-star models are consistent with their models.
The bottom panel shows that the photospheric luminosities of our models are insensitive to the metallicity and opacity parameters.
They are systematically lower than the core luminosity by $\sim 10\mbox{-}30\ \%$.
This difference corresponds to the energy used to accelerate winds.

\begin{figure}
\begin{center}
\begin{tabular}{c}
{\includegraphics[scale=0.3, angle=-90]{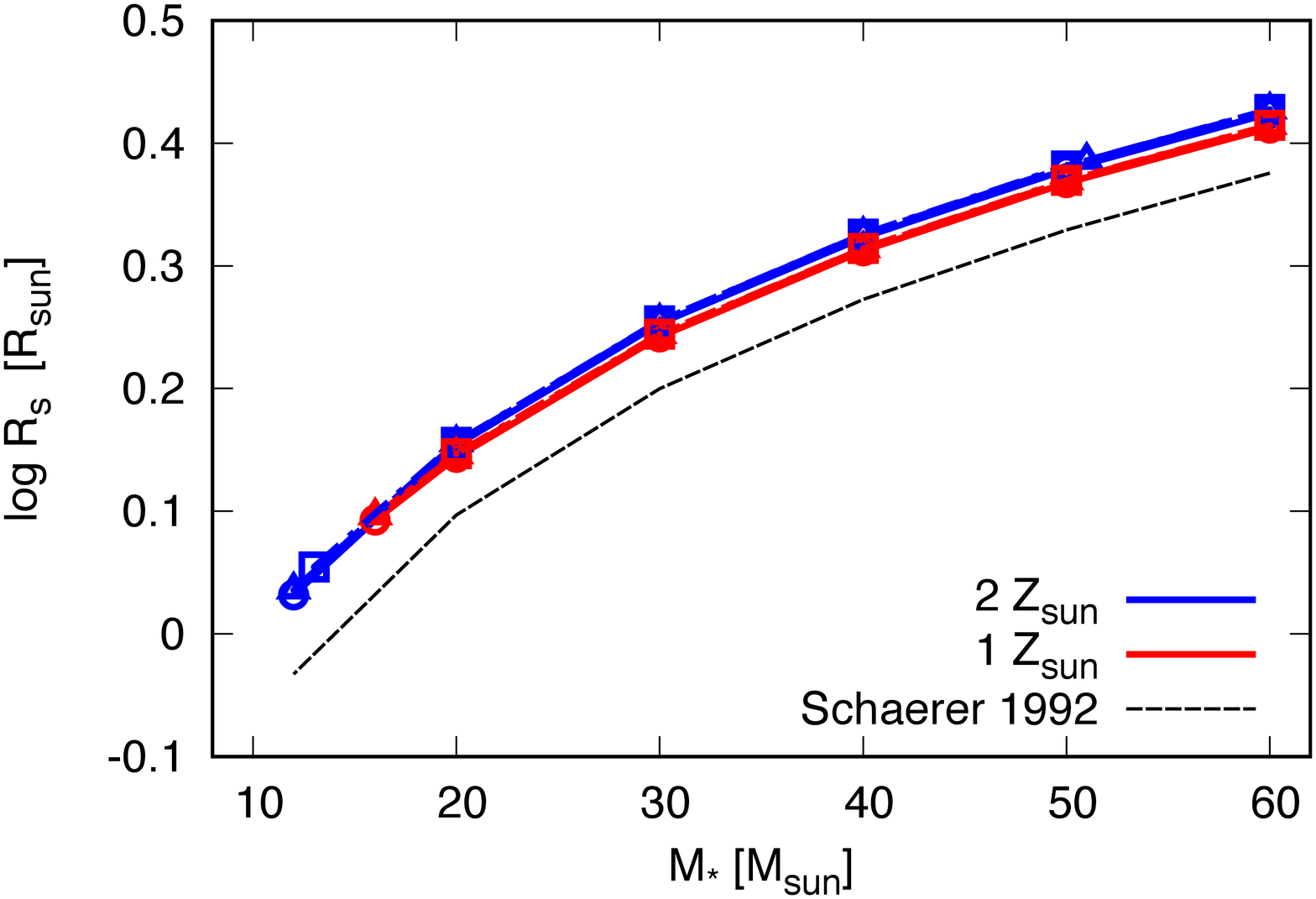}}\\
{\includegraphics[scale=0.3, angle=-90]{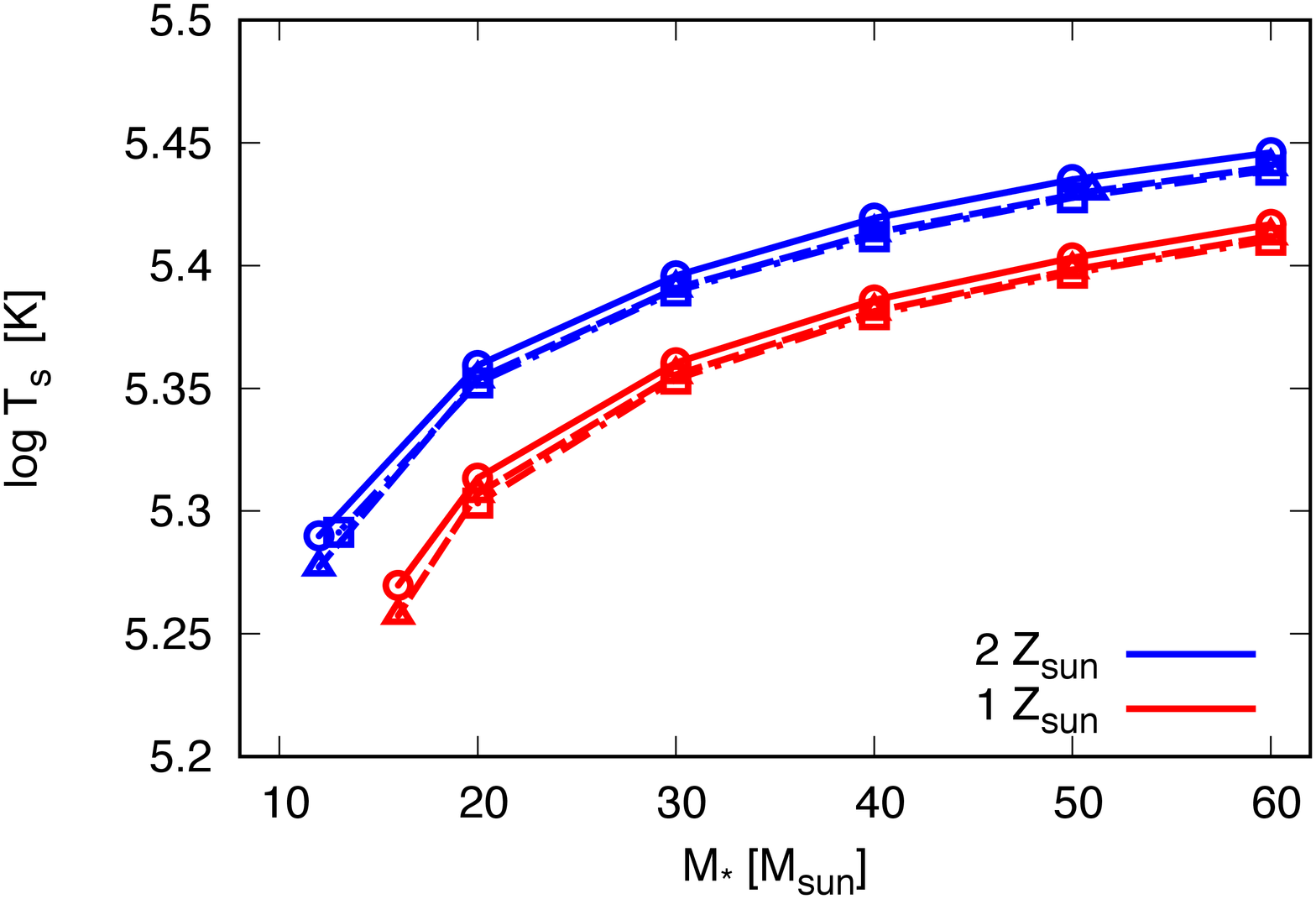}}\\
{\includegraphics[scale=0.3, angle=-90]{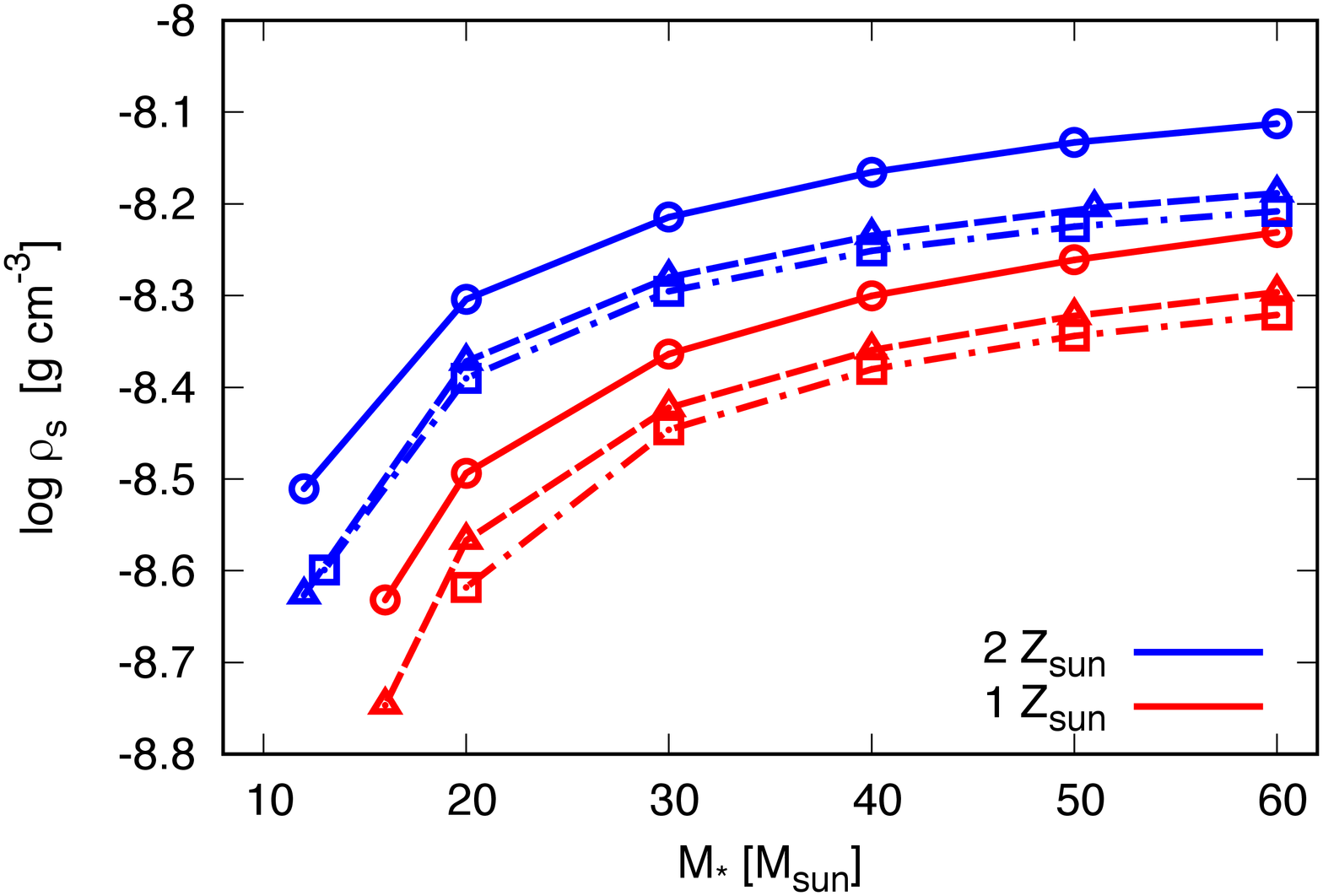}}\\
\end{tabular}
\caption{Top panel: the sonic radius as a function of the stellar mass.
The thin black dashed line shows the mass-radius relation of the hydrostatic He-star models~\citep[Eq. 4 of][]{Schaerer1992}.
Middle and bottom panels: the temperature and density at the sonic point as a function of the stellar mass.
The meanings for the symbols, line types, and colors are summarized in Table \ref{tab:parameter}.}
\label{fig:wn_sonic}
\end{center}
\end{figure}

In Fig. \ref{fig:wn_sonic}, the top panel shows the sonic radius as a function of the stellar mass.
The thin black dashed line shows the mass-radius relation of the hydrostatic He-star models~\citep[Eq. 4 of][]{Schaerer1992}.
The sonic radii hardly depend on the opacity parameters and metallicity.
Compared with the radii of \cite{Schaerer1992}'s hydrostatic models, our sonic radii are larger by about 10\ \%.

The middle and bottom panels of Fig. \ref{fig:wn_sonic} show the temperature and density at the sonic point as a function of the stellar mass.
We find that the temperature at the sonic point $T_{\rm s}$ is almost independent of the opacity parameters.
On the other hand, $\rho_{\rm s}$ weakly depends on the opacity parameters, and it takes higher values for the model with a smaller $\beta$~(compare the solid and dashed lines).
Moreover, both $T_{\rm s}$ and $\rho_{\rm s}$ become higher in more massive and/or higher metallicity models.
This can be understood from the fact that the radiation force is proportional to $\kappa_{\rm R} L_{\rm rad}$, and the luminosity and Rosseland mean opacity increases with the mass and metallicity, respectively.
The larger the radiation force is, it becomes equal to the local gravity more rapidly in the higher temperature and density regions of a wind.

\begin{figure}
\begin{center}
\begin{tabular}{c}
{\includegraphics[scale=0.3, angle=-90]{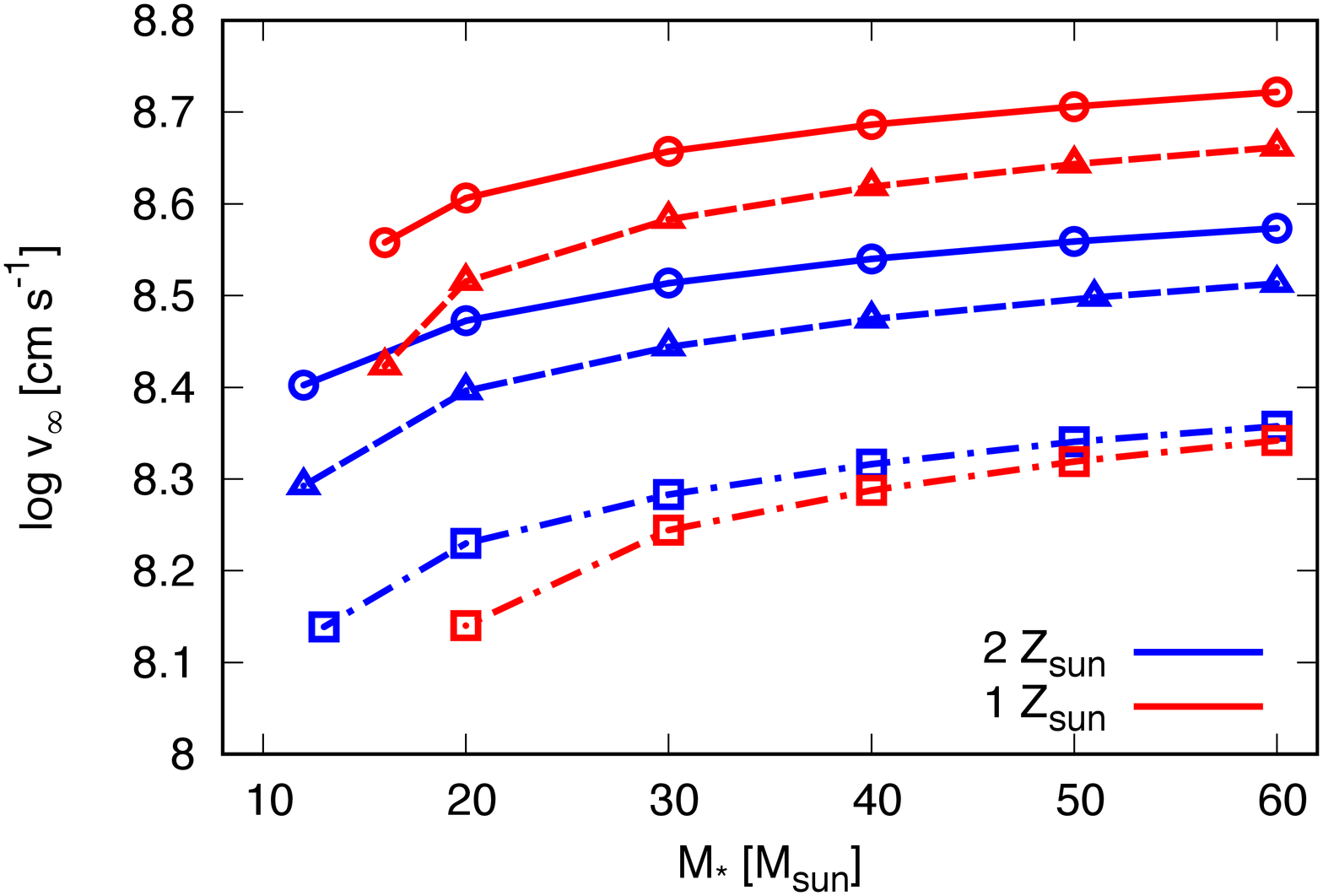}}\\
{\includegraphics[scale=0.3, angle=-90]{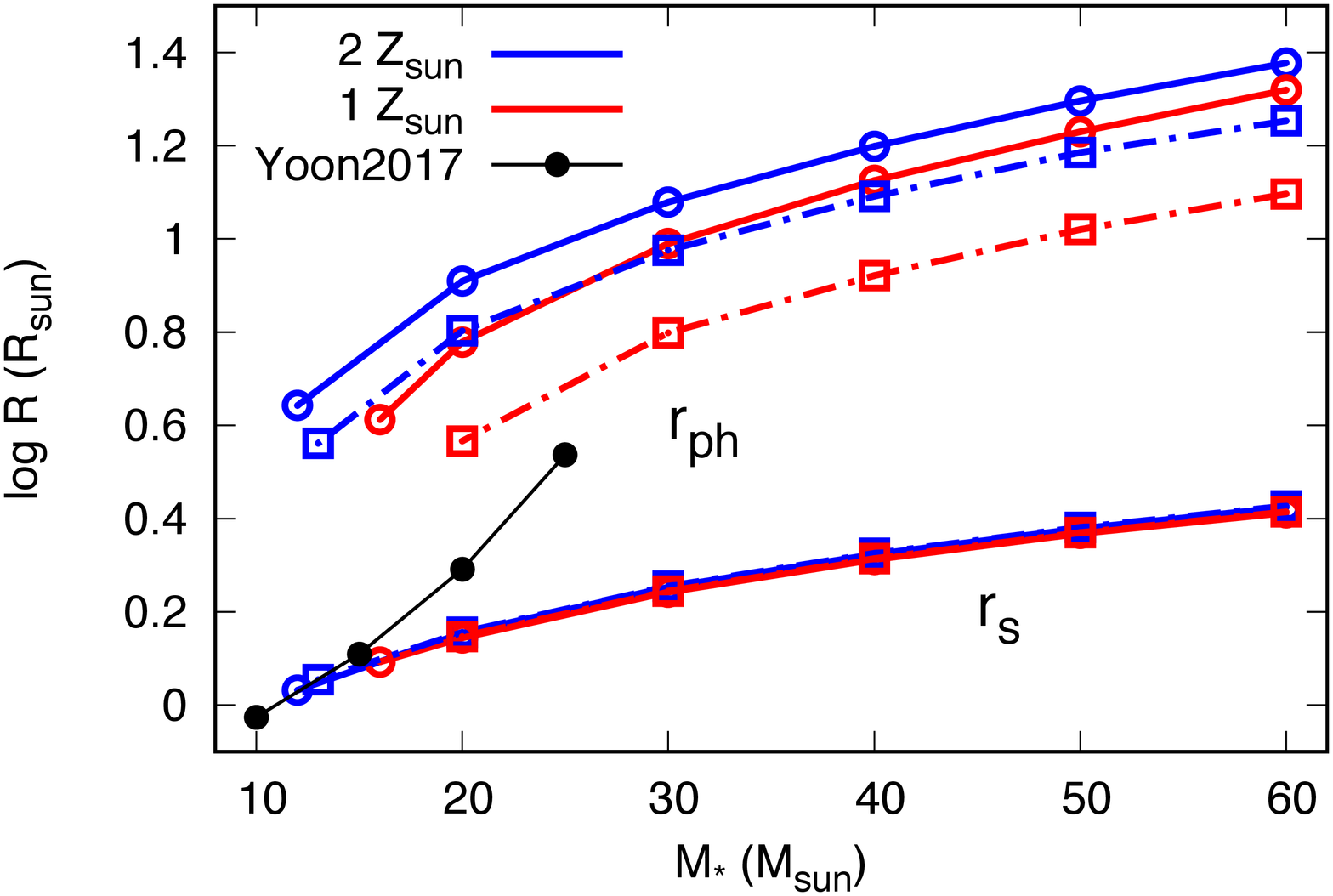}}\\
{\includegraphics[scale=0.3, angle=-90]{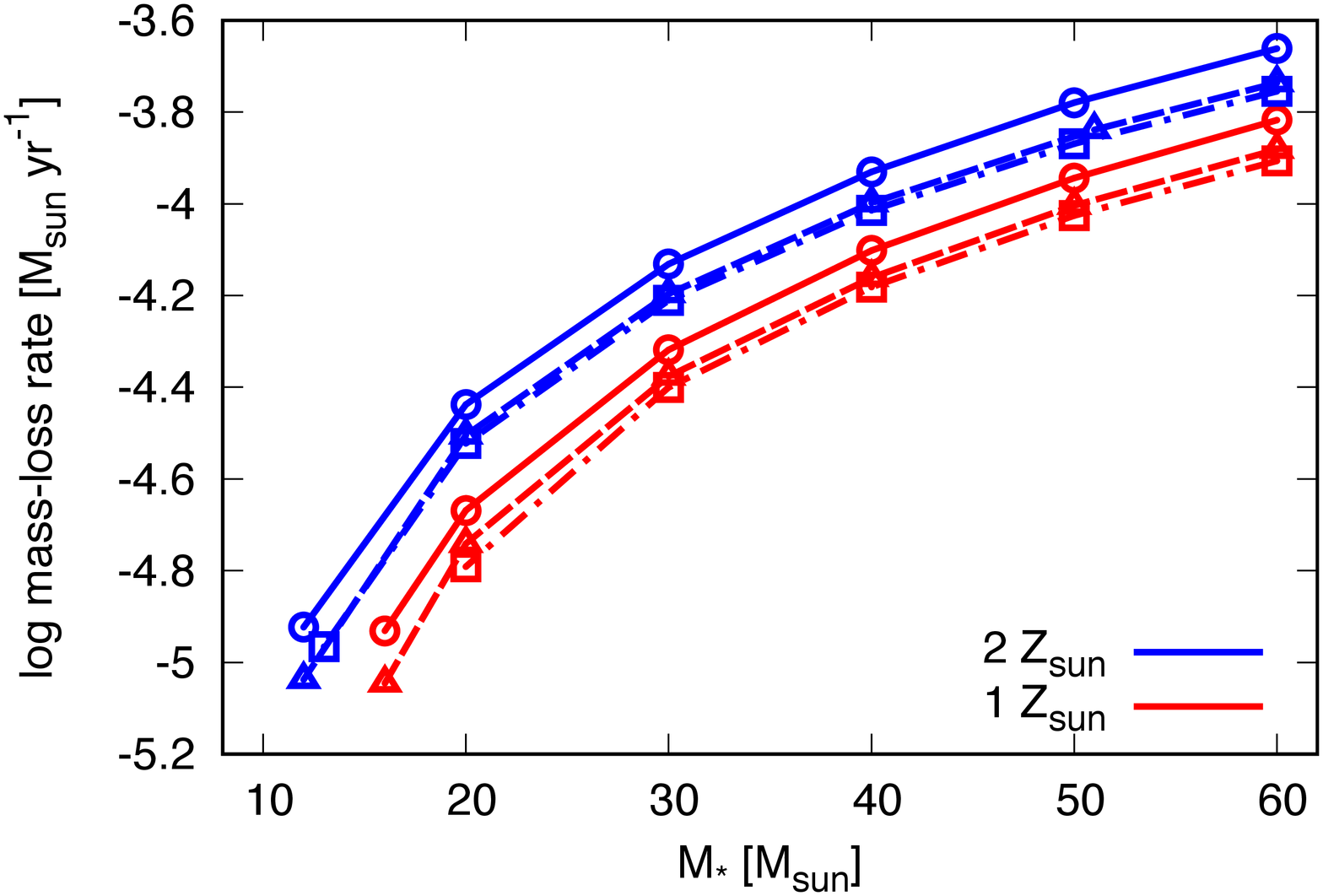}}\\
\end{tabular}
\caption{The terminal velocity~(top panel), the relation between the photospheric and sonic radii~(middle panel), and the mass-loss rate~(bottom panel) as a function of the stellar mass.
In the middle panel, the black thin line shows the photospheric radii of hydrostatic He-star models calculated by \cite{Yoon2017}.
The meanings for the symbols, line types, and colors are summarized in Table \ref{tab:parameter}.}
\label{fig:wn_wind}
\end{center}
\end{figure}

The top panel of Fig. \ref{fig:wn_wind} shows the terminal velocities as a function of stellar mass and their dependences on the metallicity and the opacity parameters.
The range of our models covers the observational values of WNE stars~\citep[$v_\infty \sim 1000\mbox{-}3000\ {\rm km}\ {\rm s}^{-1}$;][]{Hamann2006}.
The large dispersion in the terminal velocities merely reflects the difference in the opacity parameters.

The middle panel of Fig. \ref{fig:wn_wind} shows the radius at the photosphere~($r_{\rm ph}$) and at the sonic point~($r_{\rm s}$) as a function of stellar mass.
In every model, the photosphere is located at a radius by a factor of 3-10 larger than the sonic radius, indicating that the supersonic region is much more extended compared to the subsonic and static core regions.
The black thin line with filled circles shows the photospheric radii of hydrostatic He-star models calculated by \cite{Yoon2017}.
For $M_\ast \leq 15\ {\rm M}_\odot$, the photospheric radius of the static model is comparable to the sonic radius of our hydrodynamical model, but it increases more rapidly than the sonic radius for more massive models, because of the effect of the Fe-group opacity peak~(note that sonic point is always located below the opacity peak).
The photospheric radius of a static model is smaller than that of a hydrodynamic model at least for $M_\ast \leq 25\ {\rm M}_\odot$.

The bottom panel of Fig. \ref{fig:wn_wind} shows the mass-loss rate as a function of stellar mass.
Note that mass-loss rates are calculated as an eigenvalue of the equations for the given stellar mass, metallicity, and opacity parameters.
Just like $r_{\rm s}$, $T_{\rm s}$, and $\rho_{\rm s}$, mass-loss rates increase with mass and metallicity.
Mass-loss rates depend only weakly on the opacity parameters, and the dispersion in the mass-loss rates comes mainly from that of $\rho_{\rm s}$.
The mass-loss rates of our models cover the range of observational values for WNE stars, $\dot{M}_{\rm w} = 10^{-5}\mbox{-}10^{-4}\ {\rm M}_\odot\ {\rm yr}^{-1}$~\citep{Hamann2006}.

\subsection{CO-enriched Models}

\begin{figure*}
\begin{center}
\includegraphics[scale=0.6, angle=-90]{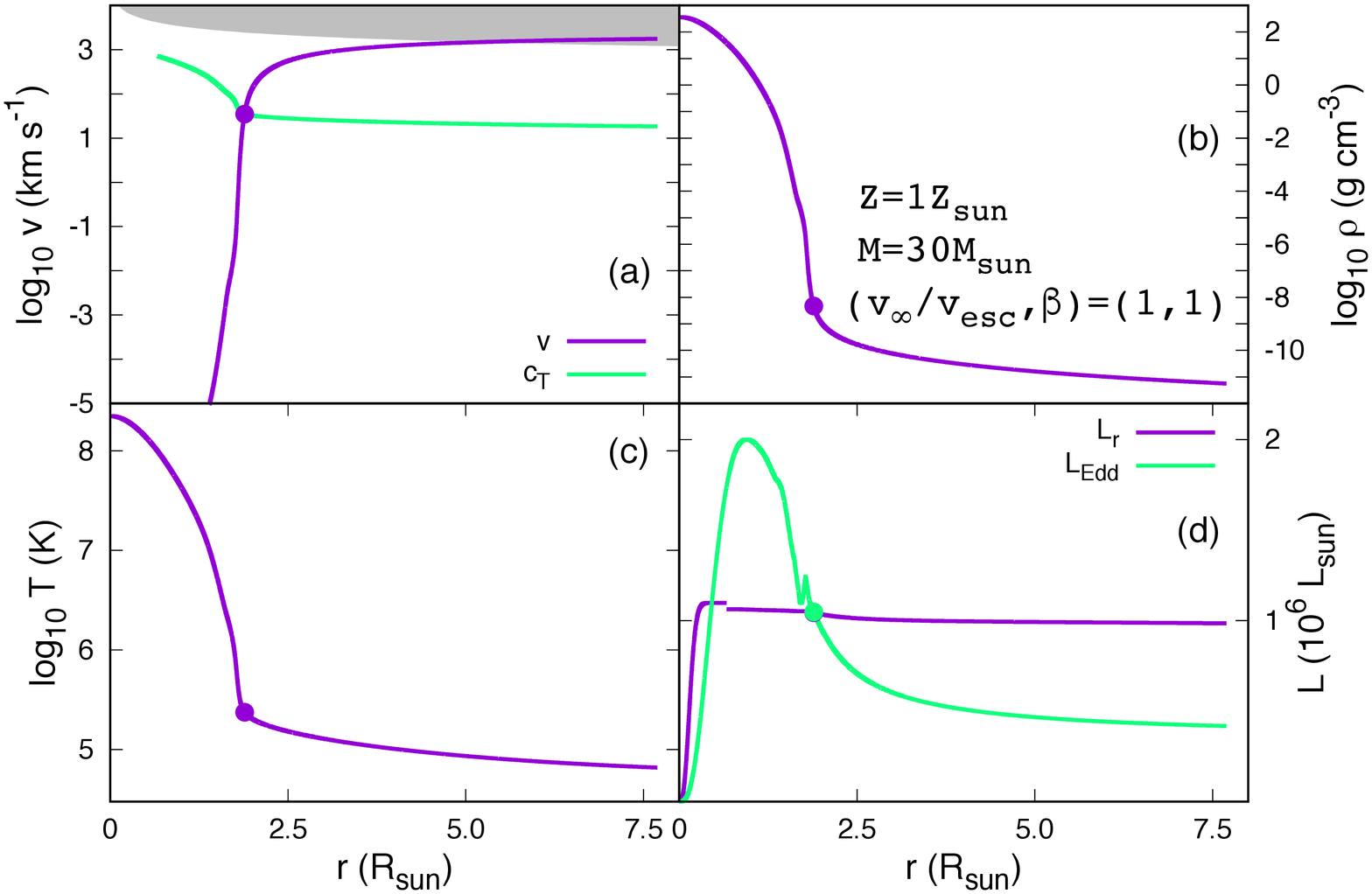}
\caption{Same as Fig. \ref{fig:wn_m30_z1_a20_b10}, but for the CO-enriched model.
The mass-loss rate of this model is $\simeq 5.7 \times 10^{-5}\ {\rm M}_\odot\ {\rm yr}^{-1}$.}
\label{fig:wc_m30_z1_a20_b10}
\end{center}
\end{figure*}

Here, we present the results of the CO-enriched models~($dX_{\rm C} = 0.4$ and $dX_{\rm O} = 0.1$).
With the larger mass fractions of C and O, the opacity bump at $\sim 2 \times 10^6\ {\rm K}$ that is produced by the bound-free transition of C and O, is larger~\citep{Iglesias1996}.
This opacity bump may help to accelerate the deeper layers of a wind.

Fig. \ref{fig:wc_m30_z1_a20_b10} shows the structure of a CO-enriched model with $M_\ast = 30\ {\rm M}_\odot$, $Z = 1\ {\rm Z}_\odot$, and $(v_\infty/v_{\rm esc}(r_{\rm s}), \beta) = (1.0, 1.0)$.
Despite the large C/O enrichment, the stellar structure is similar to the He-rich model of the same mass
in Fig. \ref{fig:wn_m30_z1_a20_b10}.
The subsonic part starts at $r_{\rm m} \simeq 0.664\ {\rm R}_\odot$, the sonic point~(filled circle) occurs at $r_{\rm s} \simeq 1.89\ {\rm R}_\odot$, and the photosphere appears at $r_{\rm ph} \simeq 7.70\ {\rm R}_\odot$.
Also in this case, the photospheric radius $r_{\rm ph}$ is several times larger than the sonic radius $r_{\rm s}$.
The temperature at the sonic point is $T_{\rm s} \simeq 2.38 \times 10^5\ {\rm K}$.
These values are slightly larger than those of the He-rich model.
The wind acceleration continues until the velocity reaches $v_\infty \simeq 1760\ {\rm km}\ {\rm s}^{-1}$, which exceeds the local escape velocity at the photosphere.
We find that the C/O opacity bump at $\sim 2 \times 10^6\ {\rm K}$ contribute only slightly to the wind acceleration.
In this model, the core luminosity, photospheric luminosity, and mass-loss rate are obtained as $L_{\rm core} \simeq 1.10 \times 10^6\ {\rm L}_\odot$, $L_{\rm ph} \simeq 9.85 \times 10^5\ {\rm L}_\odot$, and $\dot{M}_{\rm w} \simeq 5.65 \times 10^{-5}\ {\rm M}_\odot\ {\rm yr}^{-1}$, respectively.
All of them are slightly larger than those of the He-rich model, while the fraction of the radiative luminosity used for the wind acceleration is almost the same~($\simeq 10\ \%$).

In order to see the parameter dependences, we have calculated CO-enriched models with various sets of parameters for a mass range of $M_\ast = 10\mbox{-}40\ {\rm M}_\odot$.
We find that the dependences of stellar structures on the model parameters are similar to the cases of He-rich models.
The results are discussed in more detail in Appendix.

In Table \ref{tab:comp_wc}, we compare the hydrodynamic wind models for WC stars calculated by \cite{Grafener2005} with one of our CO-enriched models.
We find that the core or sonic radius, the temperature at the sonic point, and the mass-loss rate show a good agreement with each other.

\begin{table}
\caption{Comparison of the WC wind model of \cite{Grafener2005}~(GH05) and one of our CO-enriched models.}
\begin{center}
{\begin{tabular}{lcc}
\hline
                                                                                 & GH05 WC & CO-enriched \\
\hline
$M_\ast~({\rm M}_\odot)$                                        & 13.65             & 15.0     \\
$L_\ast~({\rm L}_\odot)$                                          & $10^{5.45}$  & $10^{5.56}$     \\
$R_\ast$ or $r_{\rm s}~({\rm R}_\odot)$                  &  0.905           &  1.27    \\
$T_\ast$ or $T_{\rm eff}(r_{\rm s})$~(kK)                &  140               & 127   \\
$T_{\rm s}$~(kK)                                                     &  199               &   193 \\
$\dot{M}_{\rm w}~({\rm M}_\odot\ {\rm yr}^{-1})$     & $10^{-5.14}$ & $10^{-4.93}$      \\
$v_\infty$~(${\rm km}\ {\rm s}^{-1}$)                        &  2010            &  1165       \\
 \hline
\end{tabular}}
\end{center}
{\bf Notes.} In the GH05 model, $M_\ast$, $L_\ast$, $R_\ast$, and $T_\ast$ are given as the model parameters.
In our model, $M_\ast$, $v_\infty/v_{\rm esc}(r_{\rm s})=1.0$ and $\beta=1.0$ are provided.
\label{tab:comp_wc}
\end{table}

\section{Comparison with Observation}\label{sec:obs}
In this section, we compare the results of our He-star models with the observed properties of WR stars.
He-rich and CO-enriched models are compared with WNE and WC stars, respectively.
We focus on the relation between the mass-loss rate and luminosity~(Sections \ref{subsec:wr_mdot_lum} and \ref{subsec:mdot_scale}), the ratios of radiation to gas pressure at the sonic point~(Section \ref{subsec:wr_pres}), and the positions in the HR diagram~(Section \ref{subsec:wr_hr}).

\subsection{Mass-Loss Rate Versus Luminosity}\label{subsec:wr_mdot_lum}

\begin{figure}
\begin{center}
\begin{tabular}{c}
{\includegraphics[scale=0.3, angle=-90]{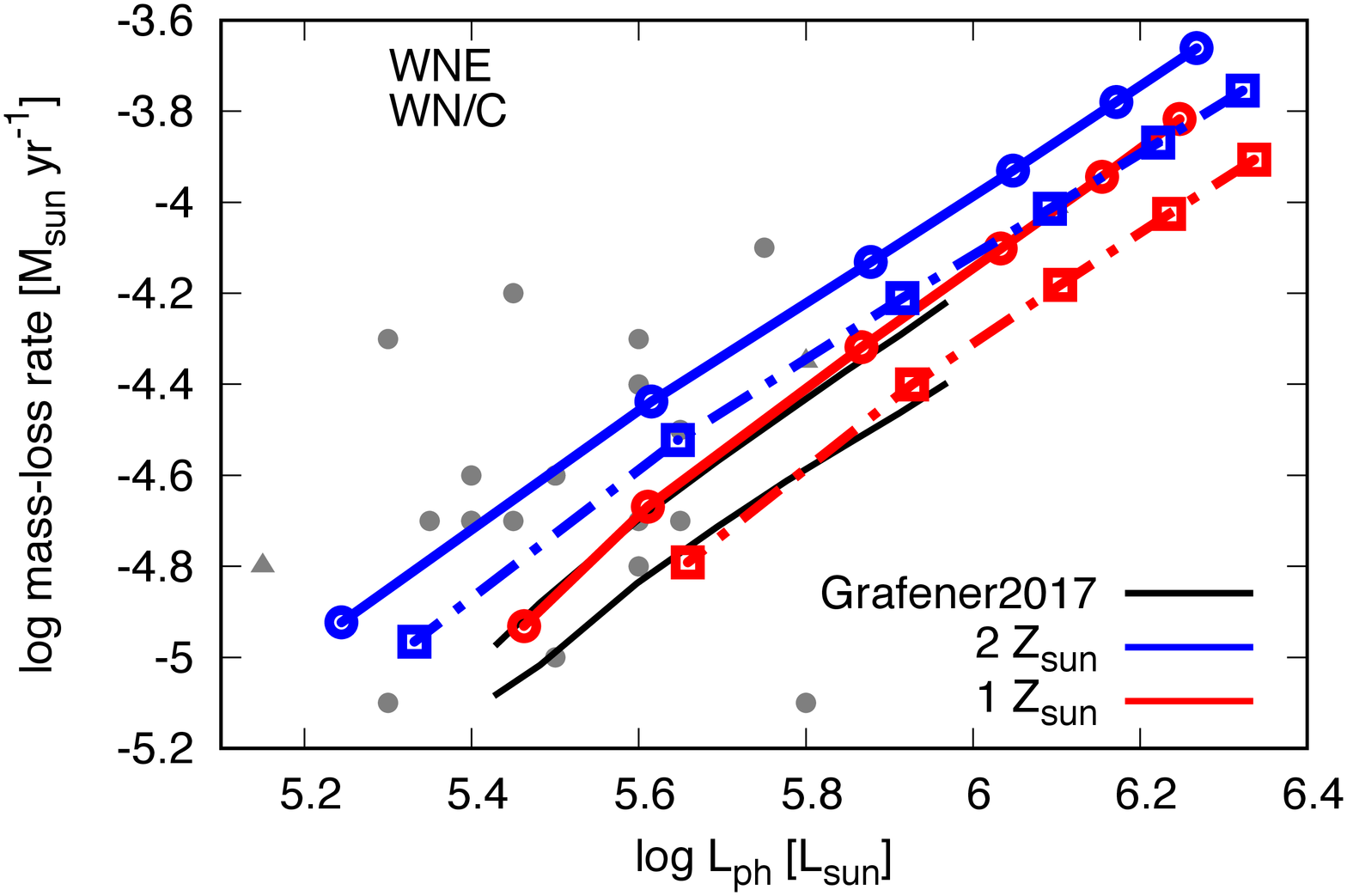}}\\
{\includegraphics[scale=0.3, angle=-90]{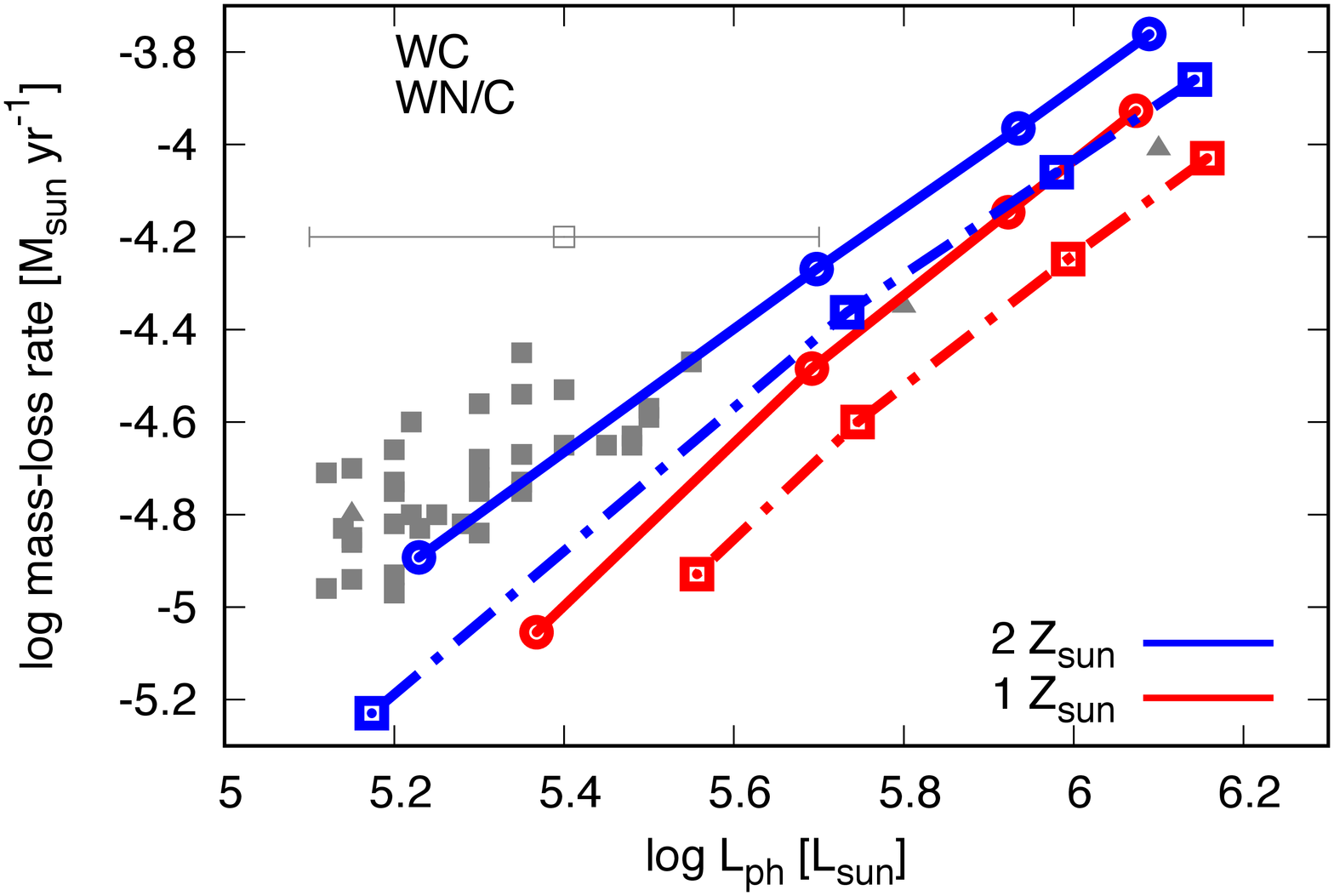}}\\
\end{tabular}
\caption{Mass-loss rates plotted as a function of photospheric luminosities.
The meanings for the symbols, line types, and colors are summarized in Table \ref{tab:parameter}.
Top panel: for the He-rich models. The filled grey circles and triangles show the observed values of the Galactic WNE stars and WN/C transition types, respectively~\citep{Hamann2006, Sander2012}.
The black solid lines show the mass-loss rates derived theoretically by \cite{Grafener2017}.
Bottom panel: for the CO-enriched models. The filled grey squares and triangles show the observed values of Galactic WC stars and WN/C transition types~\citep{Sander2012}. 
The error bar shows a typical uncertainty of the observed luminosities.}
\label{fig:wr_mdot_lum}
\end{center}
\end{figure}

In Fig. \ref{fig:wr_mdot_lum}, we show the relations between the mass-loss rate and luminosity for He-rich models~(top panel) and CO-enriched models~(bottom panel).
Here, the filled grey circles, filled grey squares, and triangles correspond to Galactic WNE stars, Galactic WC stars, and WN/C transition types, respectively~\citep{Hamann2006, Sander2012}.

Both He-rich and CO-enriched models have mass-loss rates that are comparable to the observed WR stars, as discussed in Fig. \ref{fig:wn_wind}.
Moreover, CO-enriched models have slightly larger rates compared to He-rich models for a given luminosity or mass.
This is consistent with the observations, where WC stars have larger mass-loss rates than WNE stars for a given luminosity~\citep{Yoon2017}.

Recently, \cite{Grafener2017} derived the mass-loss rates for WNE stars by matching $P_{\rm rad}$ and $P_{\rm gas}$~(or $T$ and $\rho$) at the sonic point which their hydrostatic stellar models have with those $\beta$-type wind models predict.
In the top panel, the black solid lines show the relations derived theoretically by \cite{Grafener2017}.
For the solar metallicity case, they obtained the rates of $\dot{M}_{\rm w} \approx 10^{-5.1}\mbox{-}10^{-4.2}\ {\rm M}_\odot\ {\rm yr}^{-1}$ and the scaling relation of $\dot{M}_{\rm w} \propto L_\ast^{1.3}$ for $14\mbox{-}30\ {\rm M}_\odot$.
From Fig. \ref{fig:wr_mdot_lum}, we find out the scaling relations of $\dot{M}_{\rm w} \propto L_\ast^{1.2-1.3}$ for $M_\ast \geq 20\ {\rm M}_\odot$ in both He-rich and CO-enriched models.
The rates as well as scaling relations of \cite{Grafener2017} agree quite well with those of our He-rich models.

However, theoretical mass-loss rates of both ours and \cite{Grafener2017}'s deviate from the observed rates of some WR stars.
For He-rich models, although solar metallicity models are consistent with some of the WNE stars with lower mass-loss rates, even  $2\ {\rm Z}_\odot$ models cannot explain the very high mass-loss rates of less luminous WNE stars.
For CO-enriched models, even $2\ {\rm Z}_\odot$ models show large deviation from the distribution of WC stars, although the observed values might have considerable uncertainties~(cf. the error bar in the bottom panel).
Future {\it Gaia} data releases will improve luminosity measurements and our understanding of these relationships~\citep{Gaia2016a}.

\subsection{Scaling Relations for Mass-Loss Rates}\label{subsec:mdot_scale}
For the most massive models~($M_\ast \geq 30\ {\rm M}_\odot$), a scaling relation, $\dot{M}_{\rm w} \propto M_\ast^{1.2}$, can be derived from the following simple analytical arguments.
Mass-loss rates are determined from the sonic point quantities as $\dot{M}_{\rm w} \equiv 4 \pi r_{\rm s}^2 \rho_{\rm s} v_{\rm s}$ with $v_{\rm s} \propto T_{\rm s}^{1/2}$.
Owing to the strong temperature dependence of the Fe opacity bump, $T_{\rm s}$ is almost independent of the stellar mass: $T_{\rm s} \approx 2 \times 10^{5}\ {\rm K}$.
A sonic point appears at a point where the luminosity becomes equal to the local Eddington luminosity: 
$L_\ast \approx 4 \pi c G M_\ast / \kappa_{\rm s}$ with $\kappa_{\rm s} \equiv \kappa_{\rm R}(\rho_{\rm s}, T_{\rm s})$.
Since the opacity is roughly proportional to the density around the Fe bump, $\kappa_{\rm s} \propto \rho_{\rm s}$, $\rho_{\rm s}$ depends on the stellar mass and luminosity as $\rho_{\rm s} \propto M_\ast/L_\ast$.
For the most massive models, the stellar luminosity is roughly proportional to the stellar mass: $L_\ast \propto M_\ast$~\citep[e.g.,][]{Kippenhahn2012}.
Therefore, $\rho_{\rm s}$ hardly depends on $M_\ast$ as well.
Finally, the sonic radius follows the mass-radius relation of a hydrostatic He-star model: $r_{\rm s} \propto M_\ast^{0.6}$~(Fig. \ref{fig:wn_sonic}, top panel).
Summarizing the above results leads to the simple scaling relation: $\dot{M}_{\rm w} \propto M_\ast^{1.2} \propto L_\ast^{1.2}$.
In reality, both $T_{\rm s}$ and $\rho_{\rm s}$ somewhat depend on the stellar mass, so that the actual dependence would be a little steeper $\dot{M}_{\rm w} \propto M_\ast^{1.6}$ for given metallicity and opacity parameters~(the bottom panel of Fig. \ref{fig:wn_wind}).

For WNE stars, the empirical relation of $\dot{M}_{\rm w} \propto L_\ast^{1.18}$ is derived from the observations of LMC stars by \cite{Hainich2014}, and that of $\dot{M}_{\rm w} \propto L_\ast^{1.27}$ from Galactic stars by \cite{Nugis2000} and \cite{Yoon2017}.
On the other hand, the relation of $\dot{M}_{\rm w} \propto L_\ast^{0.8}\mbox{-}L_\ast^{0.85}$ is obtained for Galactic WC stars~\citep{Nugis2000, Sander2012, Tramper2016}.
While the deviation of our relation from that of WNE stars are relatively small, it becomes significant for WC stars. 
However, the observed distributions show large scatter in both stars, so that more sample is needed for more complete discussion.

By comparing the results between $1\ {\rm Z}_\odot$ and $2\ {\rm Z}_\odot$ of Fig. \ref{fig:wr_mdot_lum}, we obtain the metallicity dependence of the mass-loss rate as $\dot{M}_{\rm w} \propto Z^{0.6-0.8}$ for $M_\ast \geq 20\ {\rm M}_\odot$ both in He-rich and CO-enriched models.
\cite{Grafener2017} derived the dependence of $\dot{M}_{\rm w} \propto Z^{0.8-1.0}$ for their WNE models of $M_\ast \geq 25\ {\rm M}_\odot$ by comparing the results between the Galactic~(Z = 0.02) and LMC metallicity~(Z = 0.008).
\cite{Vink2005} also showed theoretically the metallicity dependence of $\dot{M}_{\rm w} \propto Z^{0.86}$ for their wind models of $10^{-3}\mbox{-}1\ {\rm Z}_\odot$. 
Hence, our metallicity dependence is slightly shallower compared to other theoretical studies.
On the other hand, \cite{Yoon2017} suggested empirical relations of $\dot{M}_{\rm w} \propto Z^{0.6}$ for WNE stars, while \cite{Tramper2016} suggested $\dot{M}_{\rm w} \propto Z^{0.25}$ for WC/WO stars.
Our dependence is consistent with the empirical one for WNE case, but is slightly steeper for WC/WO case.

\subsection{$P_{\rm rad}/P_{\rm gas}$ at the Sonic Point}\label{subsec:wr_pres}
\begin{figure}
\begin{center}
\begin{tabular}{c}
{\includegraphics[scale=0.3, angle=-90]{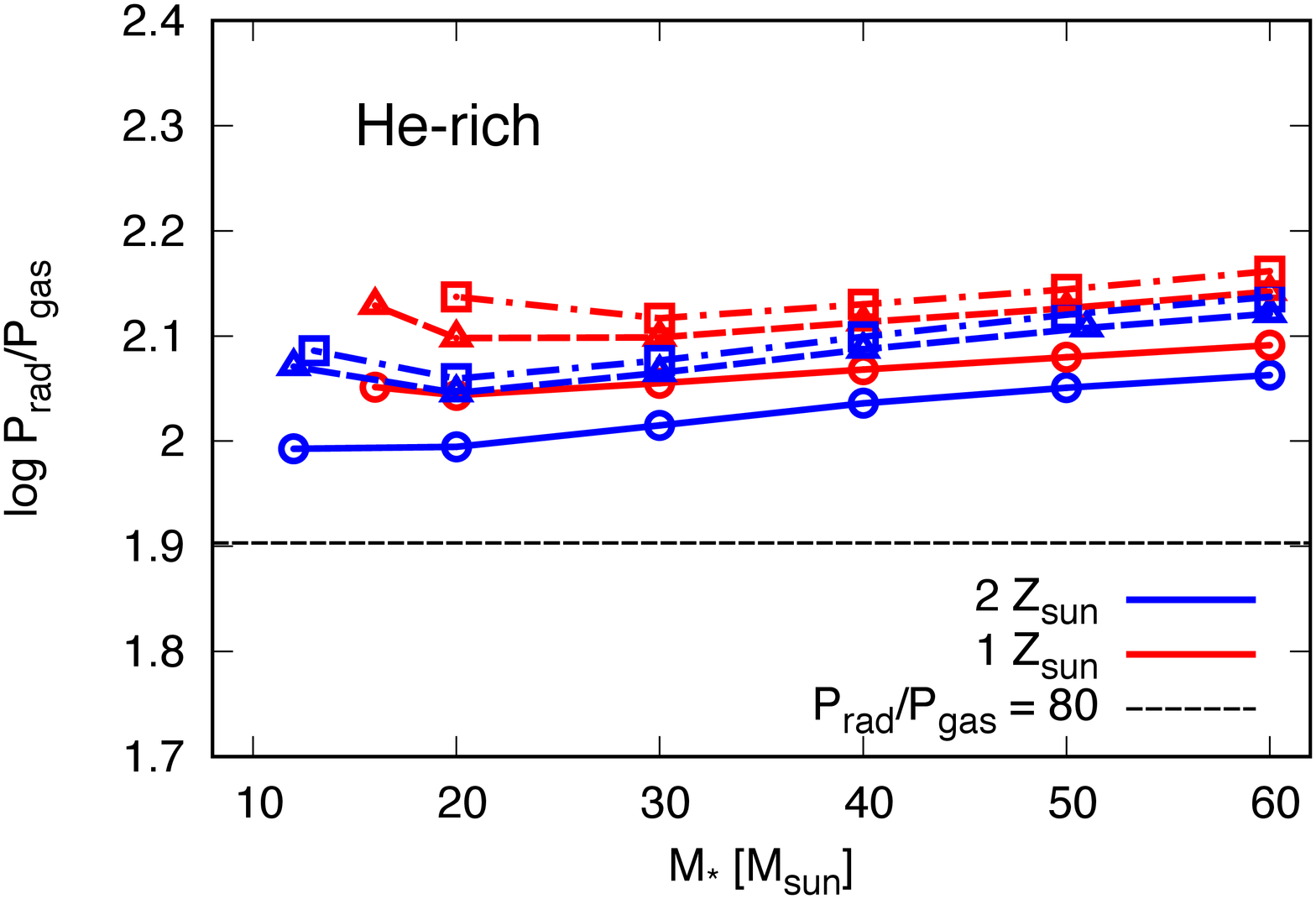}}\\
{\includegraphics[scale=0.3, angle=-90]{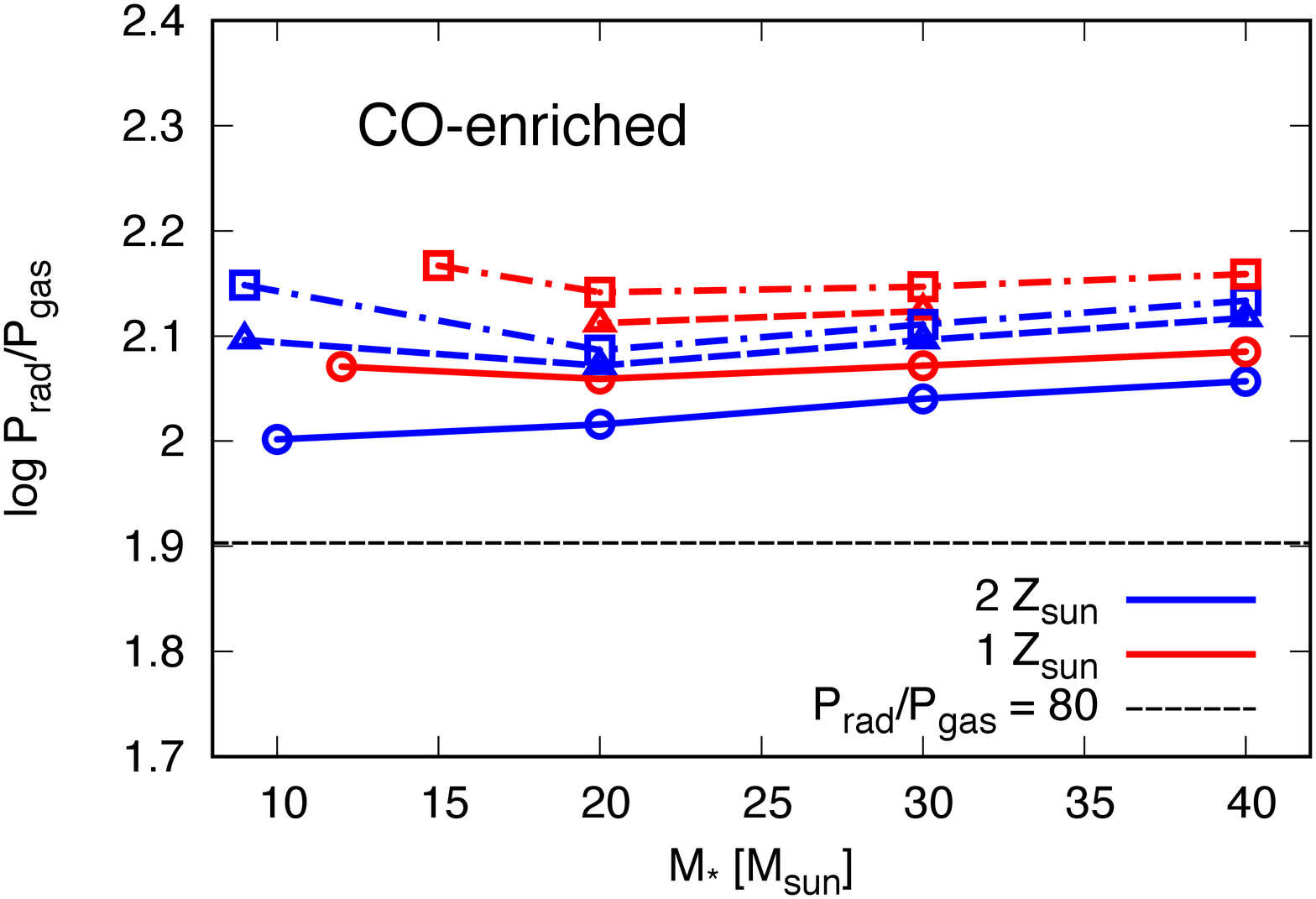}}\\
\end{tabular}
\caption{The ratios of radiation to gas pressure at the sonic point as a function of the stellar mass.
The meanings for the symbols, line types, and colors are summarized in Table \ref{tab:parameter}.
Top panel: for the He-rich models.
Bottom panel: for the CO-enriched models.}
\label{fig:wr_pres}
\end{center}
\end{figure}

\cite{Grafener2013} obtained the ratios $P_{\rm rad}/P_{\rm gas}$ at the sonic point by deriving the temperature and density there from the observed values of $L_\ast, \dot{M}_{\rm w}, v_\infty$, and $R_\ast$, and the assumed beta-type velocity profile with $\beta = 1$.
They found the typical value to be $P_{\rm rad}/P_{\rm gas} \approx 80$ for both WC and WO stars, and suggested that this value can be used as the boundary condition at the sonic point.

To compare their results with our models, we plot, in Fig. \ref{fig:wr_pres}, the ratios $P_{\rm rad}/P_{\rm gas}$ at the sonic point $r_{\rm s}$ as a function of the stellar mass for the He-rich models~(top panel) and for the CO-enriched models~(bottom panel).
The ratios lie in the narrow range of $P_{\rm rad}/P_{\rm gas} \approx 100-160$ for both models.
For a given metallicity, they tend to be smaller in the models with larger mass-loss rates~(cf. the bottom panel of Fig. \ref{fig:wn_wind}).
This is because a larger density at the sonic point leads to a larger mass-loss rate and a smaller ratio $P_{\rm rad}/P_{\rm gas}$.

In accordance with \cite{Grafener2013}, the ratios are nearly independent of mass, although the values are slightly larger than theirs.
The difference may be attributed to the fact that our models tend to have smaller mass-loss rates than the observed WR stars for a given luminosity~(Fig. \ref{fig:wr_mdot_lum}).

More recently, \cite{Grafener2017} derived for their WNE star models the ratios $P_{\rm rad}/P_{\rm gas}$ at the sonic point to be $P_{\rm rad}/P_{\rm gas} \approx 100-160$, which agree quite well with our He-rich models.

\subsection{HR Diagram}\label{subsec:wr_hr}
\begin{figure*}
\begin{center}
\begin{tabular}{c}
{\includegraphics[scale=0.45, angle=-90]{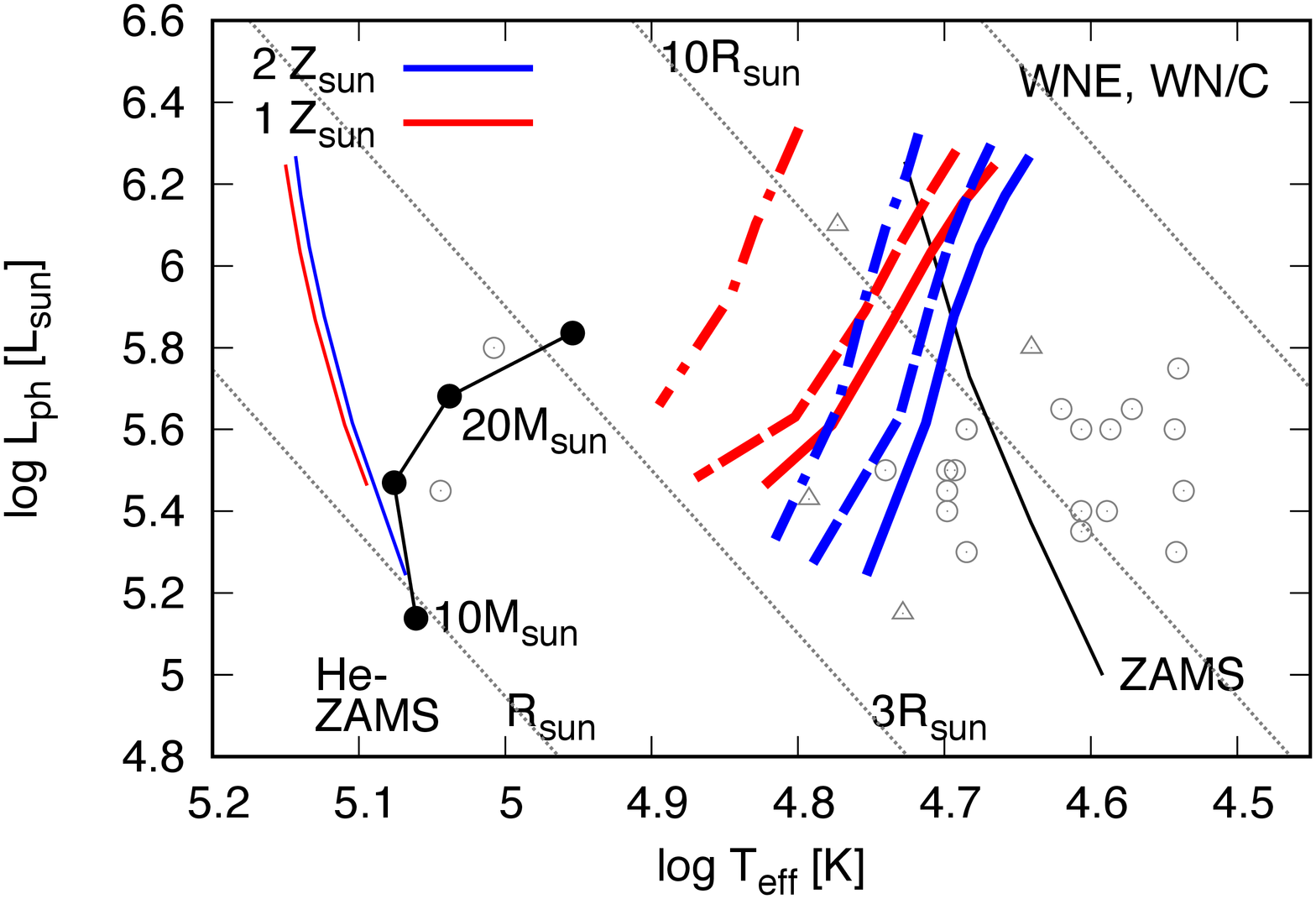}}\\
{\includegraphics[scale=0.45, angle=-90]{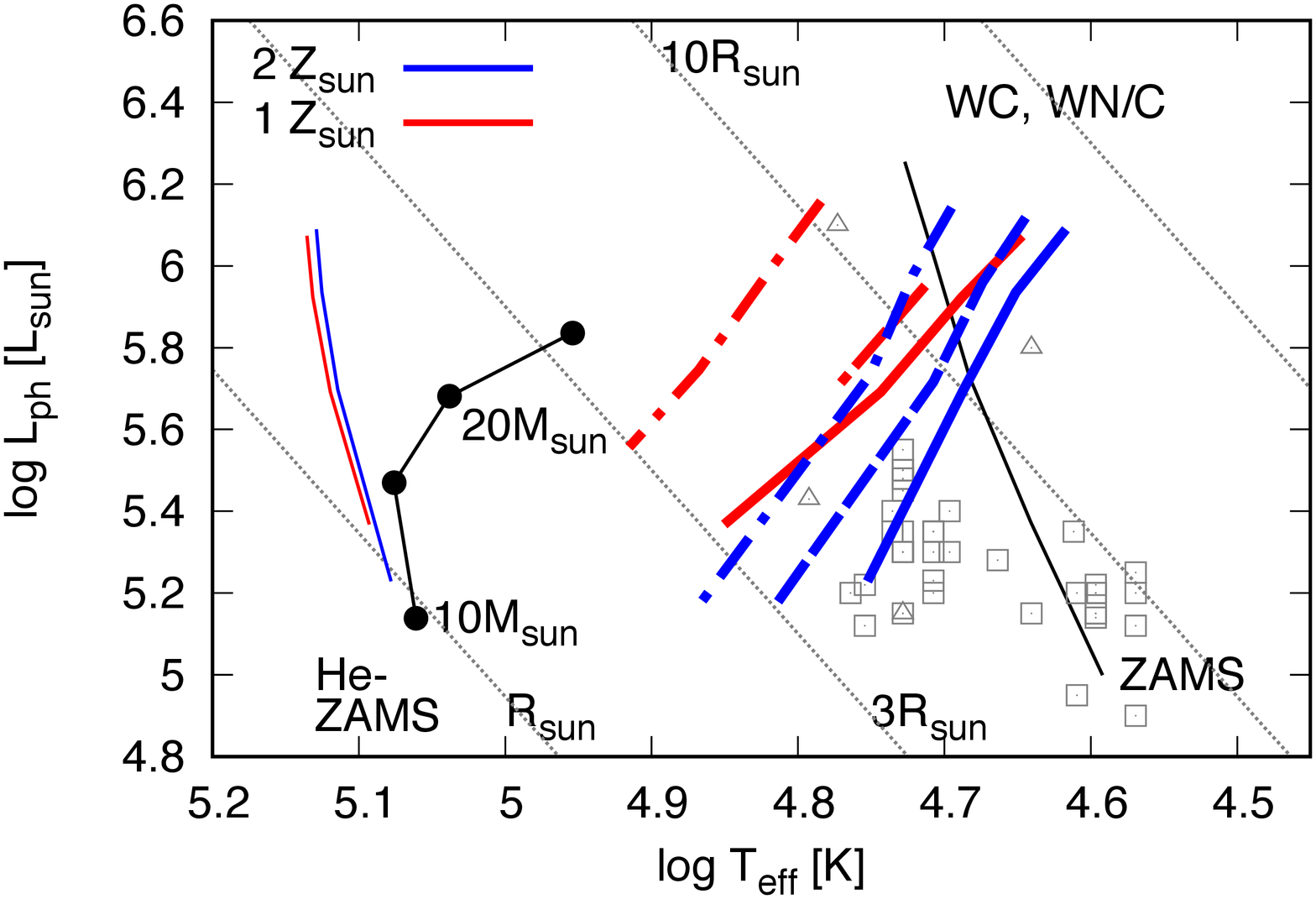}}\\
\end{tabular}
\caption{HR diagrams for the He-rich models~(top panel) and for the CO-enriched models~(bottom panel).
The thick and thin colored lines correspond to the photospheric temperature $T_{\rm ph}$ and the ``effective temperature" $T_{\rm eff}(r_{\rm s})$ evaluated at the sonic radius $r_{\rm s}$, respectively.
The red and blue lines indicate the metallicities of ${\rm Z}_\odot$ and $2\ {\rm Z}_\odot$, respectively.
The meanings for the symbols and line types are summarized in Table \ref{tab:parameter}.
Each thin black line shows the locations of the ZAMS~(right; from \cite{Schaller1992}) and He-ZAMS~(left, with filled points; from \cite{Yoon2017}).
Radius is constant along each oblique dotted line.
The open symbols indicate the ``photospheric temperatures" $T_{2/3}$ evaluated at a radius satisfying $\tau_{\rm R}(R_{2/3}) = 2/3$.
The grey circles, squares, and triangles show the locations of the Galactic WNE stars, WC stars, and WN/C transition types, respectively~\citep{Hamann2006, Sander2012}.
}
\label{fig:wr_hrd}
\end{center}
\end{figure*}

The HR diagrams in Fig. \ref{fig:wr_hrd} show the He-rich models~(top panel) and the CO-enriched models~(bottom panel).
The thick and thin colored lines correspond to the photospheric temperature $T_{\rm ph}$ and the ``effective temperature" $T_{\rm eff}(r_{\rm s})$ evaluated at the sonic radius $r_{\rm s}$, respectively.
The red and blue lines indicate the metallicities of ${\rm Z}_\odot$ and $2\ {\rm Z}_\odot$, respectively.
Each thin black line shows the locations of the ZAMS~(right; from \cite{Schaller1992}) and He-ZAMS~(left, with filled points; from \cite{Yoon2017}).
Radius is constant along each oblique dotted line.

$T_{\rm ph}$ of our He-star models are by 0.25-0.5 dex lower than $T_{\rm eff}(r_{\rm s})$, reflecting that the photospheres have 3-10 times larger radii than the sonic points~(Fig. \ref{fig:wn_wind}).
Models with more massive winds~(those with higher mass, metallicity, and $v_\infty/v_{\rm esc}(r_{\rm s})$, and smaller $\beta$) tend to have lower $T_{\rm ph}$~(and larger $r_{\rm ph}$).
Our He-star models with hydrodynamical winds have much cooler $T_{\rm ph}$~(and larger $r_{\rm ph}$) than those with static envelopes~\citep{Yoon2017} at least for $M_\ast \leq 25\ {\rm M}_\odot$.

In Fig. \ref{fig:wr_hrd}, the open symbols indicate the ``photospheric temperatures" $T_{2/3}$ evaluated at a radius satisfying $\tau_{\rm R}(R_{2/3}) = 2/3$~\footnote{The value of $T_{2/3}$ is listed as a function of $T_\ast$ and $R_{\rm t}$, the so-called ``transformed radius", on the website of the Potsdam group:
\url{http://www.astro.physik.uni-potsdam.de/~wrh/PoWR/powrgrid1.php}.
For each object, the values of $T_\ast$ and $R_{\rm t}$ are shown in the tables of \cite{Hamann2006, Sander2012}.}.
The circles, squares, and triangles show the locations of the Galactic WNE stars, WC stars, and WN/C transition types, respectively~\citep{Hamann2006, Sander2012}.

For most of the observed WR winds, photospheric temperatures are cooler than those of our He-star models.
In other words, the photospheric radius of a WR star is a few times larger than our model prediction.
This may indicate the effect of evolution that is not included in our models.

\section{Discussion}\label{sec:discussion}

\subsection{High Efficiency in Mass-Loss Rate}\label{subsec:discuss_mass_loss}

In Section \ref{subsec:wr_mdot_lum}, our He-star models underestimate the mass-loss rates for WNE stars distributing in the upper left part, and for most of the WC stars~(Fig. \ref{fig:wr_mdot_lum}).
This result implies that more efficient mechanism may be working to drive the wind of a WR star.

One possible reason for the underestimation is that we adopt the Rosseland-mean opacity that is evaluated in a static medium below the sonic point, while the opacity can be enhanced in an accelerating medium, owing to the Doppler shift of the spectral lines~\citep[e.g.,][]{Castor1975}.
\cite{Nugis2002} and \cite{Ro2016} suggested that the line force enhancement can be neglected in the subsonic layers, as long as the CAK optical depth parameter 
\begin{equation}
t_{\rm CAK} \equiv \frac{\kappa_{\rm es} \rho c_{\rm T}}{dv/dr},
\label{eq:t_cak}
\end{equation}
is larger than unity.
Here, $\kappa_{\rm es}=0.2\ {\rm cm}^2\ {\rm g}^{-1}$ is the electron scattering opacity.
At the sonic point of our models, we find $t_{\rm CAK}(r_{\rm s}) \approx 1\mbox{-}6$, which increase with mass and metallicity.
This indicates that the line force enhancement is not important below the sonic points in our models.

Another possibility to gain higher efficiency for wind acceleration is a significant enhancement in the opacity of the Fe-group elements.
Recent laboratory measurements for the opacity of the Fe-group elements suggest that the Rosseland-mean opacity near the Fe-group peak should be enhanced by 75 \% or more with respect to the values given in the OPAL table~\citep{Bailey2015, Turck-Chieze2016}.
In addition, \cite{Daszynska-Daszkiewicz2017} claimed that as large as a factor 3 increase of the opacity at $\log T = 5.47$ is needed to explain the seismic behavior of a B-type variable star.

Furthermore, the core evolution, which is not included in our models, affects the static core size and hence the radius at the sonic point. This may change the prediction of mass-loss rates considerably~(especially for WC stars).  

\subsection{Extended Photospheres of WR Stars}\label{subsec:discuss_wr_hr}

In Section \ref{subsec:wr_hr}, we find that for most of the observed WR winds, photospheric temperatures are cooler than those of our He-star models~(Fig. \ref{fig:wr_hrd}).
This result indicates that WR stars have more extended photospheres than our model predictions.

One possible reason for the deviation is that we do not consider the evolution of a hydrostatic core.
It may be important for the positions of WR stars in the HR diagram.
\citet{McClelland2016} showed that the positions of WNE stars can be explained by the evolutionary sequence of He-stars, if the mass-loss rate is smaller than empirical values and the wind clumping is large enough.
In their model, WC stars can also be explained by He-giant stars of $< 8\ {\rm M}_\odot$, if the enhanced chemical mixing occurs in the envelope.
\cite{Yoon2017} also suggested that WC stars can be explained by the evolutionary sequence of $15\ {\rm M}_\odot$ He stars, by using a new empirical mass-loss formulae.

Another possibility is that we assume a simple $\beta$-type wind in the supersonic region.
\cite{Grafener2005} constructed a steady wind model of a WC star, where the velocity structure, non-LTE level populations, and radiation field are calculated self-consistently~\citep[see][for the WN and O star cases]{Grafener2008, Sander2017}.
They found that wind acceleration does not occur continuously as expected in a single $\beta$-type law, but occurs separately in several regions, and that the velocity structure is better fitted with a double $\beta$-type law.
A sophisticated modeling of the supersonic wind would be needed to reproduce the very extended photospheres of WR stars.

\subsection{Wind Clumping}\label{subsec:clumping}
Spectroscopic observations have suggested that the winds of WR stars are inhomogeneous and clumpy~\citep{Moffat1988, Hillier1991, Hamann1998}.
In the supersonic regions of our models, clumping effects are implicitly included through the assumption of a $\beta$-type velocity law, which is equivalent to assuming an enhanced opacity in these layers~(see Eq. \ref{eq:kap_eff} and Fig. \ref{fig:wn_temp_kap}).

If the subsonic layers of a WR wind are inhomogeneous and clumpy, the mean opacity and the acceleration efficiency there could be enhanced compared to the smooth wind case~\citep{Grafener2005}.
In hydrostatic models, the clumping effect increases the inflation of the outermost envelopes both for He-ZAMS and its evolutionary models~\citep{Grafener2012, McClelland2016}.
In our He-star models with hydrodynamical envelope/wind, the effect of clumping in subsonic layers, if it exists, can be inferred by comparing the results between $1\ {\rm Z}_\odot$ and $2\ {\rm Z}_\odot$ models in Figs. \ref{fig:wn_sonic} and \ref{fig:wn_wind}, i.e., an enhanced opacity in the subsonic layers would not change the sonic radius, but increase the mass-loss rate.

While inhomogeneities in the subsonic layers can be produced by the turbulent convection caused by the Fe-group peak, their relationship with the wind-driving mechanism is still unclear~\citep{Blaes2003,Cantiello2009,Jiang2015,Grassitelli2016}.
Multidimensional and global radiation hydrodynamical simulations that cover from the wind driving region to the highly supersonic region would be needed to study this.

\section{Summary}\label{sec:summary}

We construct He-star models with optically thick winds and compare them with the properties of Galactic WR stars.
Hydrostatic He-core solutions are connected smoothly to trans-sonic wind solutions that satisfy the regularity conditions at the sonic point.
By constructing a center-to-surface structure, a mass-loss rate can be obtained as an eigenvalue of the equations.
We study how the structures in the wind launching regions and photospheres, and mass-loss rates depend on the assumed $\beta$-type velocity structures in the supersonic regions.

The rapid acceleration of the subsonic layers starts with the increase of the Fe-group opacity.
The wind velocity reaches the sound speed at $T_{\rm s} = 10^{5.25}\mbox{-}10^{5.45}\ {\rm K}$~(below the Fe-group opacity peak), where the radiation force becomes comparable to the local gravity.
Both $T_{\rm s}$ and $\rho_{\rm s}$ increase with mass and metallicity.
The mass-$r_{\rm s}$ relation is proportional to the mass-radius relation of the He-ZAMS~\citep{Schaerer1992}, but shifted upward by 10-20\ \%.
Our models have the ratios $P_{\rm rad}/P_{\rm gas}$ at the sonic point lying in the narrow range of $P_{\rm rad}/P_{\rm gas} \approx 100\mbox{-}160$.

The terminal velocities and mass-loss rates of our He-star models cover those of the observed WR winds.
The photosphere is located at a 3-10 times larger radius compared to the sonic radius.
While the terminal velocities and radius ratios $r_{\rm ph}/r_{\rm s}$ vary directly with the opacity parameters, the mass-loss rates depend only weakly on them.
Just as $r_{\rm s}$, $T_{\rm s}$, and $\rho_{\rm s}$, mass-loss rates increase with mass and metallicity.
The following relation $\dot{M}_{\rm w} \propto M_\ast^{1.2} \propto L_\ast^{1.2}$ holds for the most massive models~($M_\ast \geq 30\ {\rm M}_\odot$).

In the $\dot{M}_{\rm w}$ and $L_{\rm ph}$ plane, our models are consistent with the observations for the WNE stars with lower mass-loss rates.
Our models tend to underestimate the mass-loss rates for some WR stars.
In the HR diagram, the photospheric temperatures of observed WR winds tend to be cooler than those of our models, meaning that WR stars have more extended photospheres.
Those disagreements may be caused by the possible underestimate of the opacity around the sonic point, or various effects disregarded in this paper; such as core evolution, detailed non-LTE radiative transfer, and wind clumping.

\acknowledgments
DN would like to thank Professor Kazuyuki Omukai for continuous encouragement. 
We also thank Dr. Andreas Sander for kindly telling us the way to estimate the photospheric temperatures for the observed WR stars.
This work is supported in part by the Grant-in-Aid from the Ministry of Education, Culture, Sports, Science and Technology (MEXT) of Japan, No.16J02951~(DN).

%



\appendix
\section{Parameter Dependences of CO-enriched models}\label{subsec:WC_mdot}
\begin{figure}
\begin{center}
\begin{tabular}{cc}
\includegraphics[scale=0.3, angle=-90]{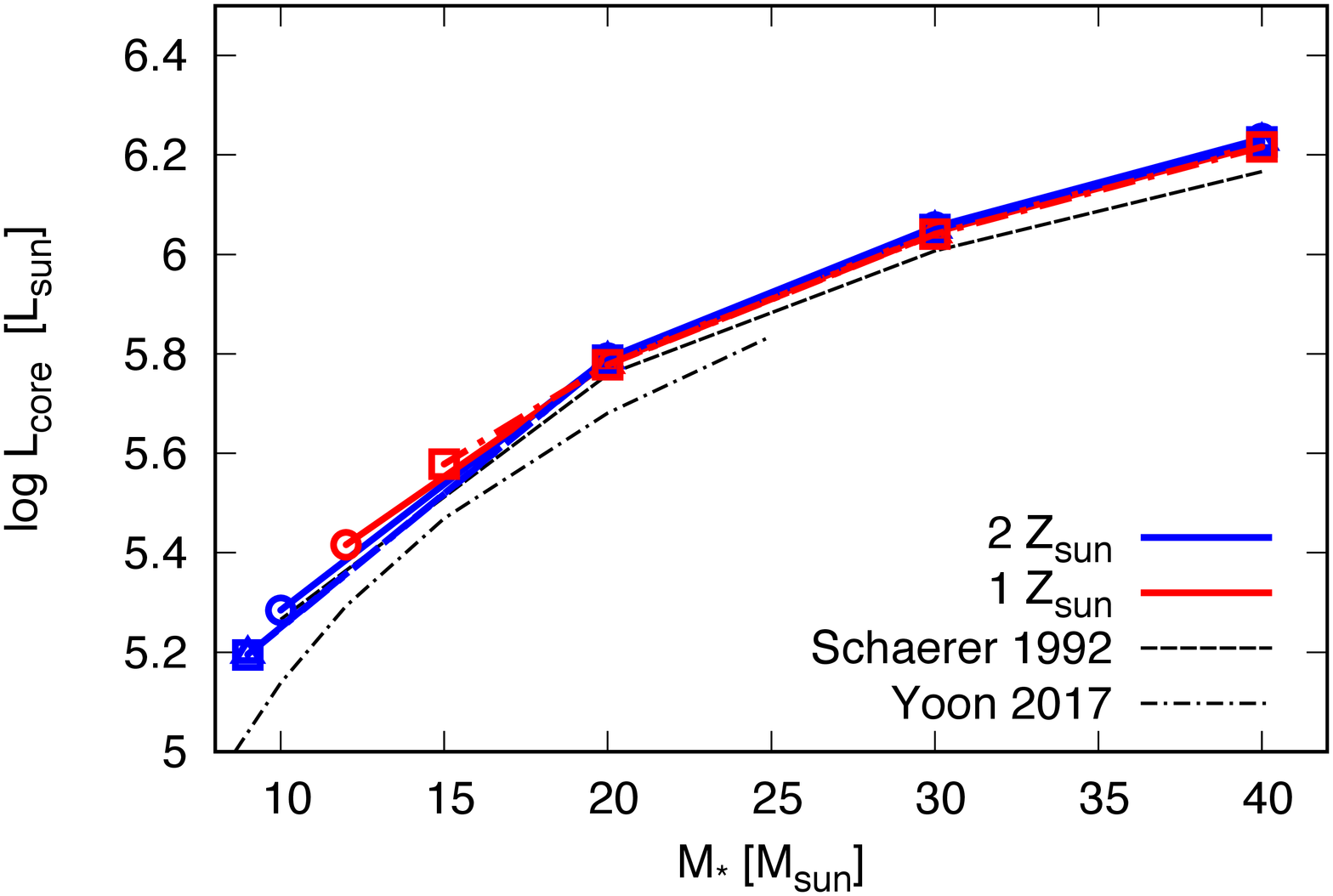}&
\includegraphics[scale=0.3, angle=-90]{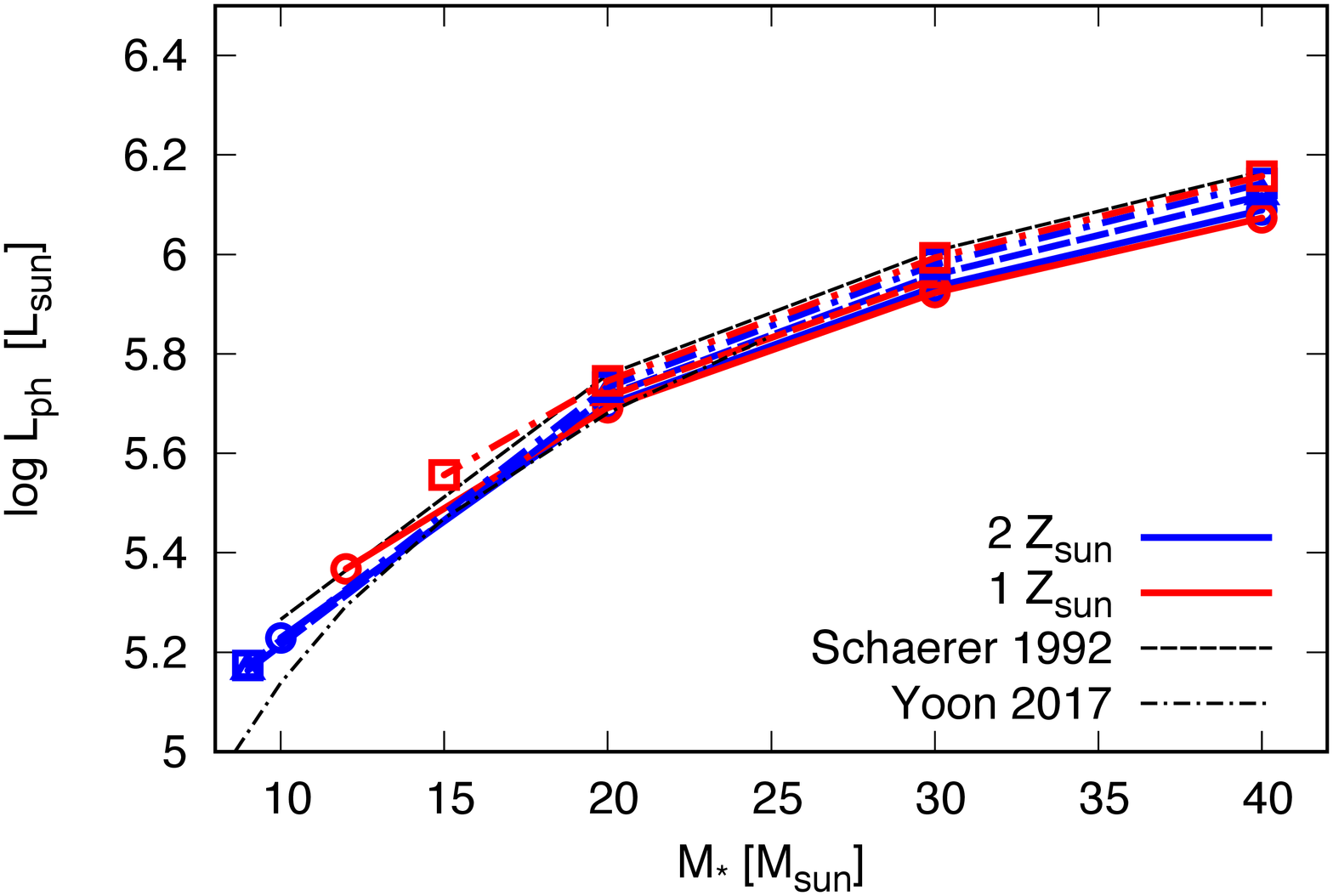}
\end{tabular}
\caption{Same as Fig. \ref{fig:wn_lum}, but for the CO-enriched models.}
\label{fig:wc_lumc}
\end{center}
\end{figure}

In Fig. \ref{fig:wc_lumc}, we show the core and photospheric luminosity as a function of the stellar mass.
CO-enriched models also follow the mass-luminosity relation of the hydrostatic He-star models, but they have slightly larger luminosities than He-rich models.
The fraction of the core luminosity used to accelerate the winds distributes among $\simeq 5\mbox{-}30\ \%$.

\begin{figure}
\begin{center}
\begin{tabular}{cc}
{\includegraphics[scale=0.3, angle=-90]{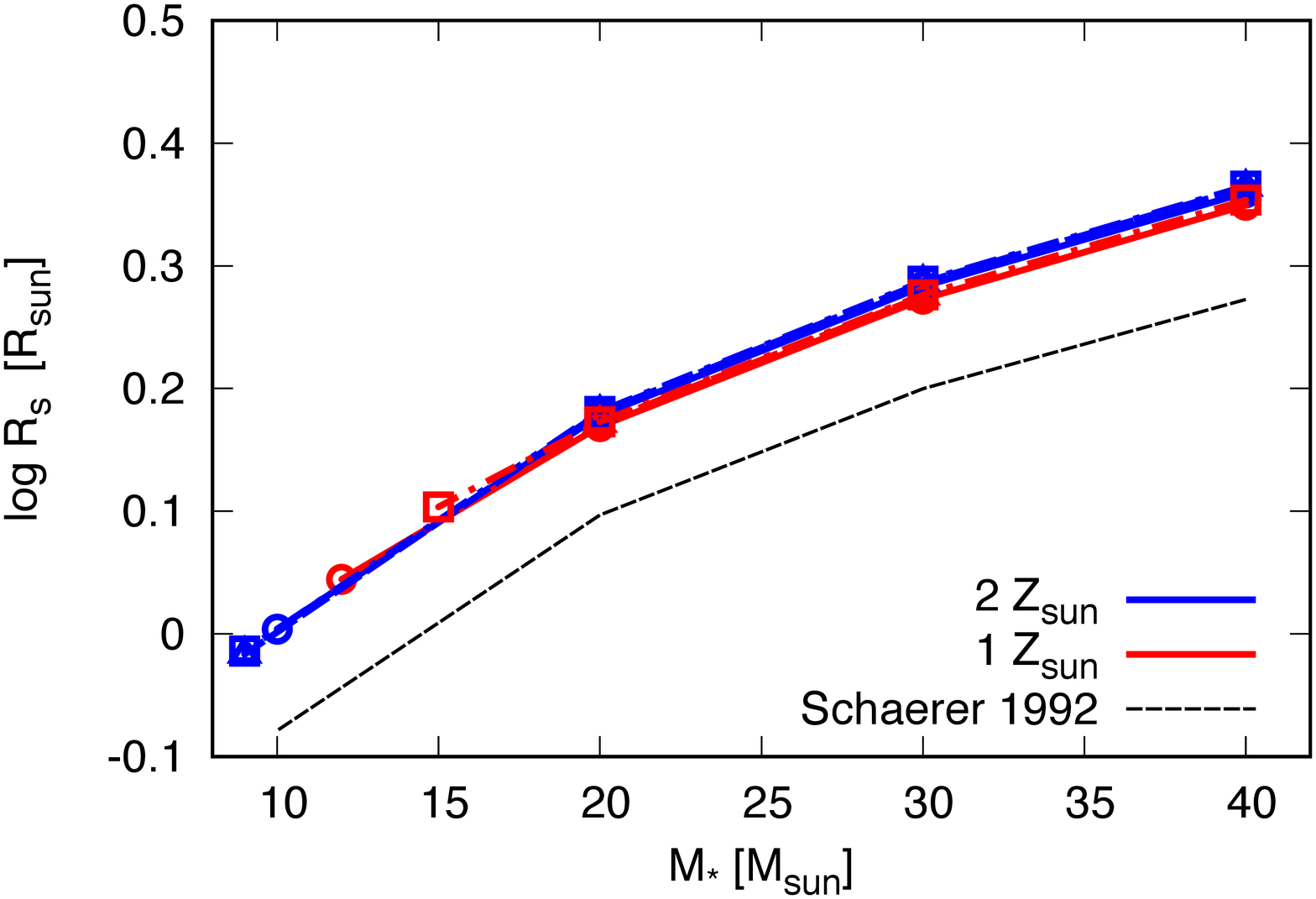}}&
{\includegraphics[scale=0.3, angle=-90]{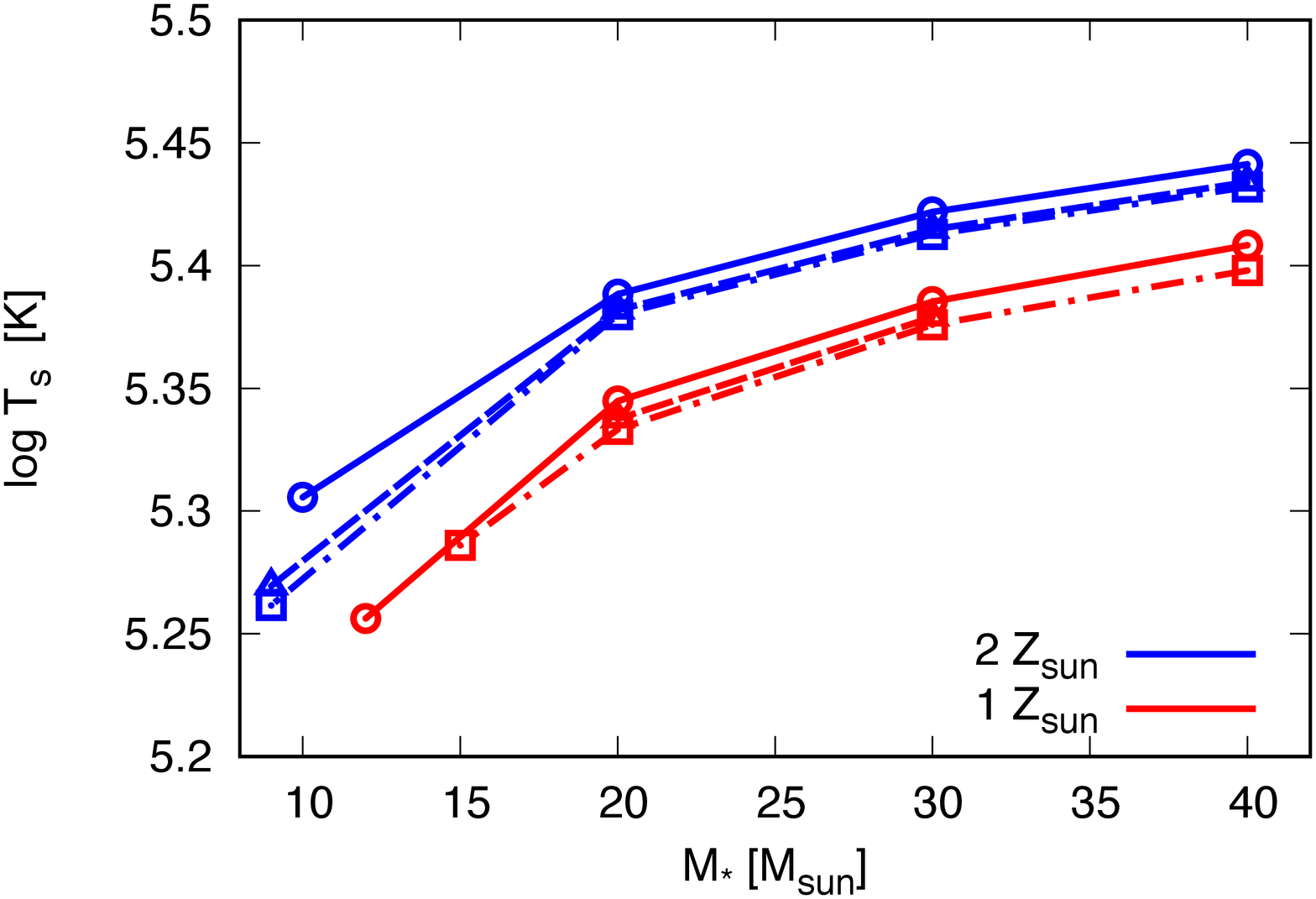}}\\
{\includegraphics[scale=0.3, angle=-90]{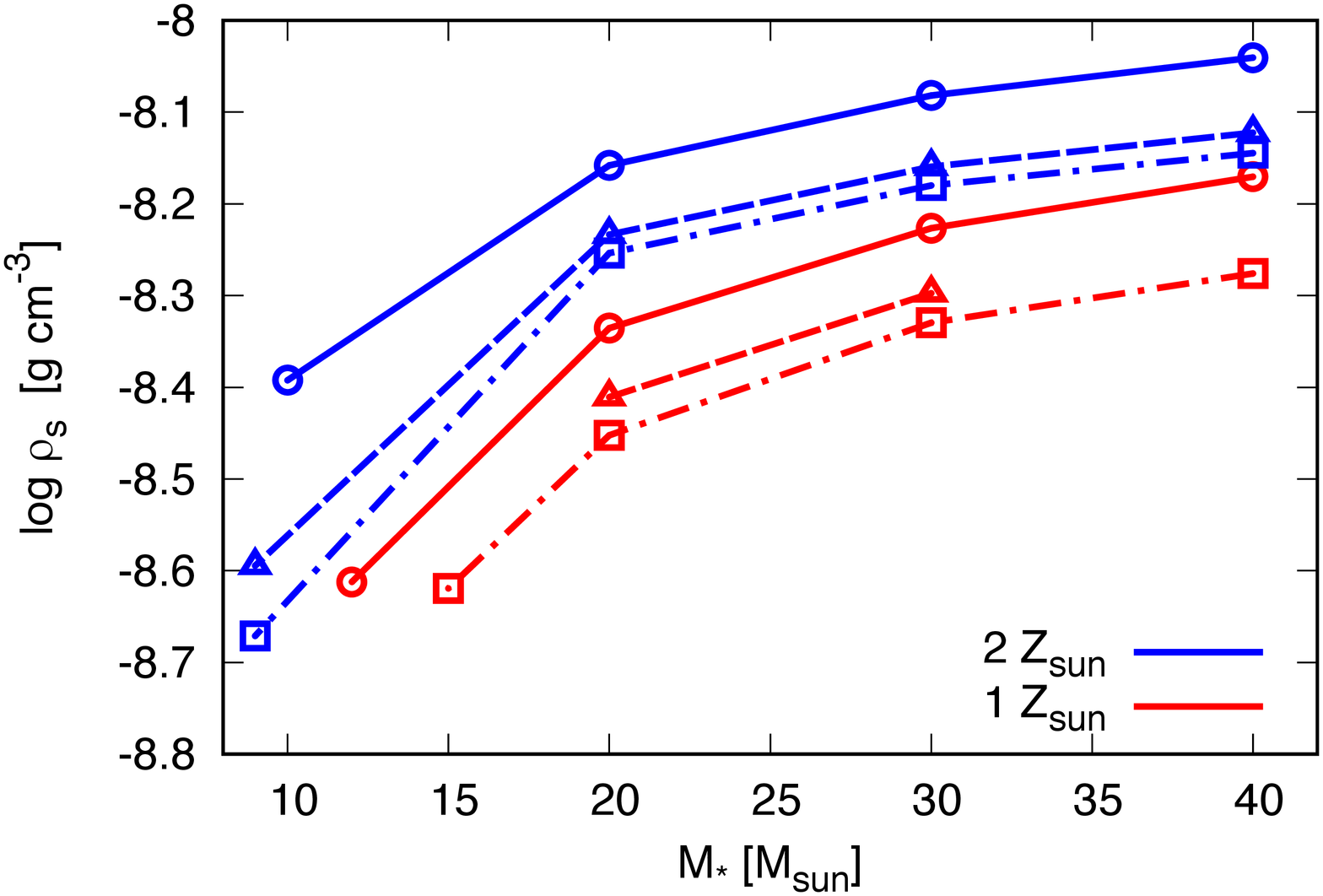}}&
{\includegraphics[scale=0.3, angle=-90]{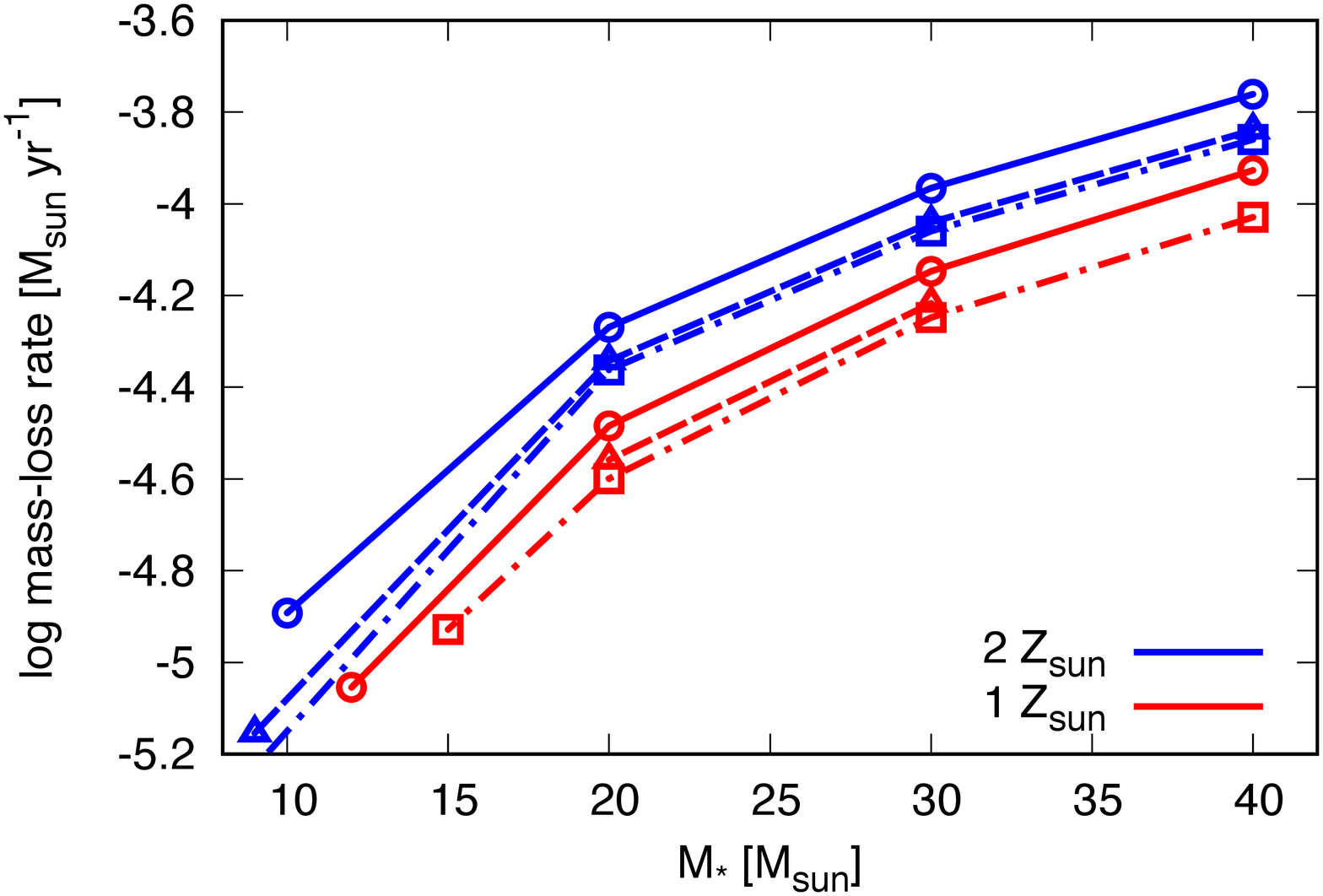}}\\
\end{tabular}
\caption{The radius~(top, left), temperature~(top, right), and density~(bottom, left) at the sonic point, and the mass-loss rate~(bottom, right) as a function of the stellar mass for the CO-enriched models.
The thin black dashed line shows the mass-radius relation of the hydrostatic He-star models~\citep[Eq. 4 of][]{Schaerer1992}.}
\label{fig:wc_sonic}
\end{center}
\end{figure}

In Fig. \ref{fig:wc_sonic}, we show the radius~(top, left), temperature~(top, right), and density~(bottom, left) at the sonic point, and the mass-loss rate~(bottom, right) as a function of the stellar mass.
While the sonic radii are larger than the radii of \cite{Schaerer1992}'s models by about 20\ \%, the subsonic layers do not show any inflation.
CO-enriched models show the similar dependences on the model parameters as He-rich models, but tend to have slightly larger values for $T_{\rm s}$, $\rho_{\rm s}$, and $\dot{M}_{\rm w}$.
This is because the radiation force approaches the local gravity more rapidly in the subsonic region, owing to the higher luminosity.
The observed WC stars have mass-loss rates of $\dot{M}_{\rm w} \simeq (1\mbox{-}4) \times 10^{-5}\ {\rm M}_\odot\ {\rm yr}^{-1}$~\citep{Sander2012, Tramper2016}, which are covered by our models.

\end{document}